\documentclass[journal]{IEEEtran}
\ifCLASSINFOpdf
\else
\fi

\usepackage{graphicx}
\usepackage{amsmath}
\usepackage{mathtools}
\usepackage{commath}
\usepackage{comment}

\usepackage{enumitem}

\usepackage{booktabs}
\usepackage{multirow} 
\usepackage{subfig}
\usepackage{epstopdf}
\usepackage{array}
\usepackage{wrapfig}

\newcommand{\innermid}{\nonscript\;\delimsize\vert\nonscript\;}
\newcommand{\activatebar}{%
  \begingroup\lccode`\~=`\|
  \lowercase{\endgroup\let~}\innermid 
  \mathcode`|=\string"8000
}

\usepackage{amsthm,amssymb}
\usepackage{setspace}
\usepackage{url}

\newcommand{\subparagraph}{}
\usepackage[compact]{titlesec}

\begin{document}


%
\title{Learning to Predict Streaming Video QoE: Distortions, Rebuffering and Memory}
%
%
%

\author{Christos~G.~Bampis,~\IEEEmembership{Student~Member,~IEEE},  
      and~Alan~C.~Bovik,~\IEEEmembership{Fellow,~IEEE}
\thanks{C. Bampis and A. C. Bovik are with the Department
of Electrical and Computer Engineering, University of Texas at Austin, Austin,
USA (e-mail: bampis@utexas.edu; bovik@ece.utexas.edu). This work is supported by Netflix Inc.}
\thanks{Manuscript received ...; revised ... .}}

%
%

%

\pdfoutput=1



\maketitle

\begin{abstract}

Mobile streaming video data accounts for a large and increasing percentage of wireless network traffic. The available bandwidths of modern wireless networks are often unstable, leading to difficulties in delivering smooth, high-quality video. Streaming service providers such as Netflix and YouTube attempt to adapt their systems to adjust in response to these bandwidth limitations by changing the video bitrate or, failing that, allowing playback interruptions (rebuffering). 

Being able to predict end users' quality of experience (QoE) resulting from these adjustments could lead to perceptually-driven network resource allocation strategies that would deliver streaming content of higher quality to clients, while being cost effective for providers. Existing objective QoE models only consider the effects on user QoE of video quality changes or playback interruptions. For streaming applications, adaptive network strategies may involve a combination of dynamic bitrate allocation along with playback interruptions when the available bandwidth reaches a very low value. 

Towards effectively predicting user QoE, we propose Video Assessment of TemporaL Artifacts and Stalls (Video ATLAS): a machine learning framework where we combine a number of QoE-related features, including objective quality features, rebuffering-aware features and memory-driven features to make QoE predictions. We evaluated our learning-based QoE prediction model on the recently designed LIVE-Netflix Video QoE Database which consists of practical playout patterns, where the videos are afflicted by both quality changes and rebuffering events, and found that it provides improved performance over state-of-the-art video quality metrics while generalizing well on different datasets. The proposed algorithm is made publicly available at \url{http://live.ece.utexas.edu/research/Quality/VideoATLAS_release_v2.rar}.

\end{abstract}

\begin{IEEEkeywords}
subjective quality of experience, video quality assessment, video streaming
\end{IEEEkeywords}

%
\IEEEpeerreviewmaketitle

\section{Introduction}
%
%
%
%
\IEEEPARstart{M}obile video traffic accounted for 55 percent of total mobile data traffic in 2015, according to the Cisco Visual Networking Index (VNI) and global mobile data traffic forecast \cite{cisco}. Since video data traffic and streaming services are significantly increasing, content providers such as Netflix and YouTube must make resource allocation decisions and mediate tradeoffs between operational costs and end user Quality of Experience (QoE). Since in video data applications such as streaming the human is the end user, perceptually-driven optimization strategies are desireable to guide the resource allocation problem.

While the motivation for perceptually-driven models is obvious, QoE prediction is still far from being an easy task. The low-level human visual system (HVS) is complex and driven by non-linear processes not yet well understood. There are also cognitive factors that influence perceived QoE, adding further layers of complexity, complicating the analysis of human subjective data and the design of QoE prediction models. For example, subjective QoE is affected by recency: more recent QoE experiences may have a higher impact on currently perceived QoE \cite{hands2001recency}. We are interested here in two types of subjective QoE: retrospective QoE and continuous-time QoE. In studies of retrospective QoE, subjects provide a single score describing their overall QoE on each presented video sequence. Studies of continuous-time QoE involve the real-time measurement of each subject's current QoE, which may be triggered by changes in video quality or streaming and by short or long term memory effects.

With respect to these challenges, we will show that existing objective video quality assessment (VQA) methods inadequately model subject QoE. There is also a broad spectrum of video distortions ranging from video compression artifacts to rebuffering events, all having different effects on subject QoE. In streaming applications, rebuffering currently appears to be a necessary evil, since the available bandwidth is volatile and hard to predict. However, only recently have sophisticated approaches been developed that predict the effects of rebuffering on QoE. Yet, making unified QoE predictions involving diverse impairments remains an elusive goal. 

Towards solving this challenging problem, we have developed a learning-based approach to making QoE predictions when the videos are afflicted by both bitrate changes and rebuffering. This is commonly seen in practice, where video bitrate often varies over time and where rebuffering events frequently occur. However, most existing subjective video quality datasets cannot be used to study general QoE models, since they either do not contain both rebuffering and quality changes, or they are of limited size or their design is not suitable for streaming applications. Towards filling this gap, the recently introduced LIVE-Netflix dataset \cite{DB_paper} was specifically designed for this problem and includes the outcomes of a large subjective study.

The rest of this paper is organized as follows. Section \ref{prev_work} discusses previous work on QoE prediction related to streaming applications. Then, Section \ref{dataset} gives an overview of the LIVE-Netflix dataset \cite{DB_paper} that we use to study these impairments and to develop more general QoE models. Section \ref{is_enough_section} investigates whether currently used VQA methods are suitable for QoE prediction on this dataset and motivates the need for a more general framework. Section \ref{our_method} describes the proposed learning-based QoE prediction framework and Section \ref{our_results} presents experimental results. Finally, Section \ref{the_end} gives conclusion.

\section{Previous work on QoE Prediction}
\label{prev_work}

QoE prediction models typically consider a set of video impairments in light of human subjective data. To facilitate a description of previous work on QoE prediction, consider the following two types of video impairments that affect perceived user QoE:

\begin{figure*}[tp]
\centerline{
\includegraphics[width=\columnwidth] {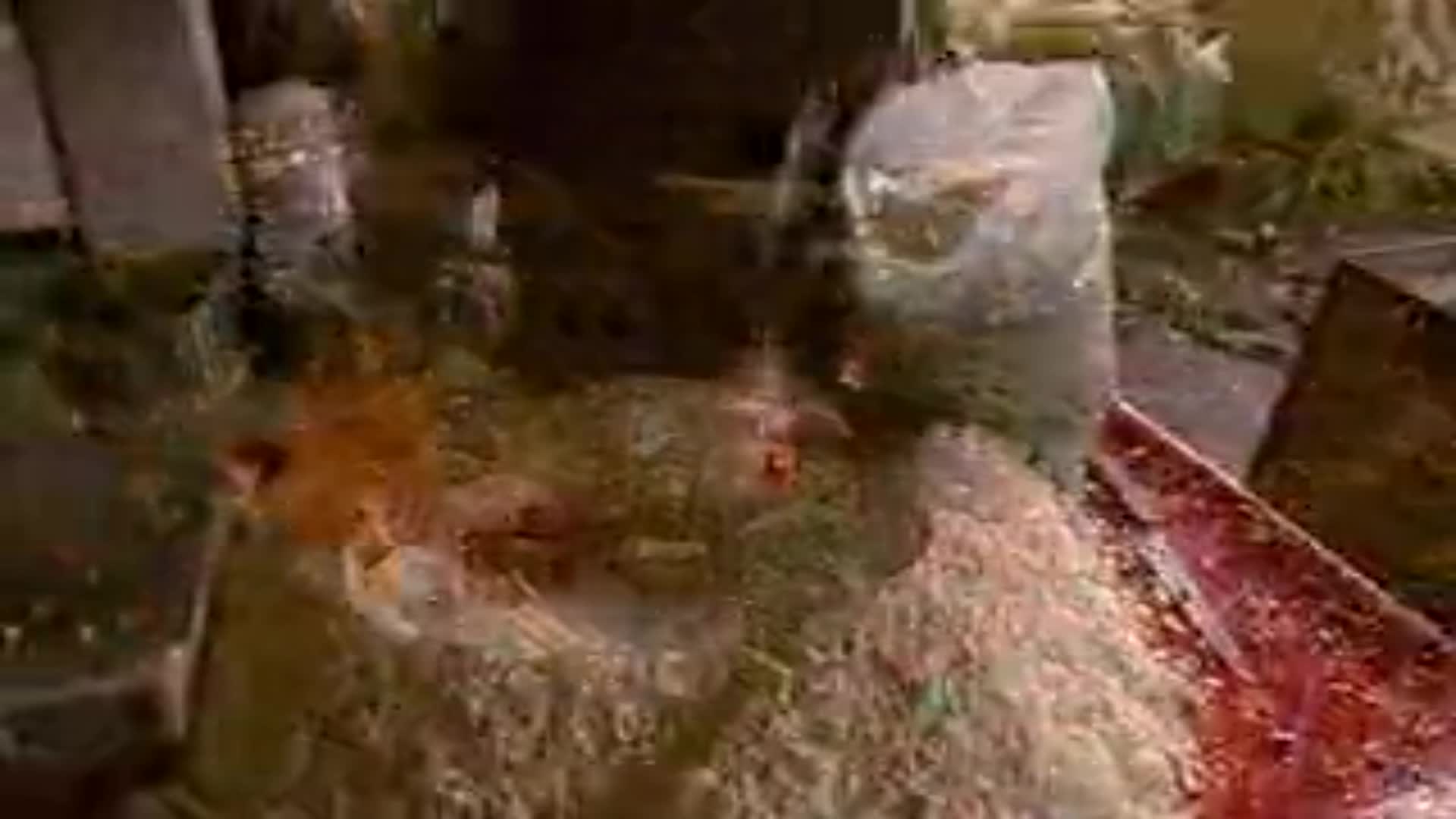}
\includegraphics[width=\columnwidth] {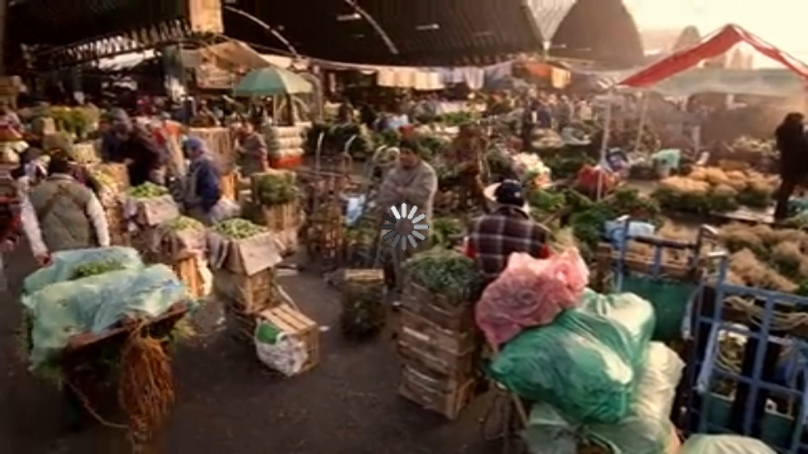}
}
\caption{Example video frames from the LIVE-Netflix dataset. Left (\ref{fig_examples}a): H.264 compression; right (\ref{fig_examples}b): playback interruption.}
\label{fig_examples}
\end{figure*}

\textit{A. Impairments of Videos with Normal Playback}

The most typical streaming scenario is to apply an adaptive bitrate allocation strategy such that bandwidth consumption is optimized. An example of a compressed video can be seen in Fig. \ref{fig_examples}a. The effects of bitrate changes on the retrospective QoE may vary according to a number of QoE-related scene aspects: low-level content (slow/fast motion scenes), previous bitrates, frequency of bitrate shifts and their noticeability, the display device being used and so on \cite{moorthy2012video}. Apart from bitrate selection schemes which lead to compression artifacts, other network-related distortions arise from packet losses \cite{liang2008analysis} or impairments of the source videos. A commonality of these impairments is that there are no implied playback interruptions, with the rare exception of severe packet loss, where whole groups of frames cannot be properly decoded. To help study and measure the video quality degradations induced by these video distortions, many successful datasets have been built \cite{5404314, moorthy2012video, de2010h, choi2015motion}. An overview of available video quality datasets can be found in \cite{winkler2012analysis}.

A wide variety of video quality assessment (VQA) models have been proposed ranging from full-reference (FR) to no-reference (NR) \cite{moorthy2011visual}. These include standard frame-based techniques (FR-IQA) such as SSIM \cite{wang2004image, wang2004video} and MS-SSIM \cite{wang2003multiscale}, temporal FR-VQA methods such as VQM\_VFD \cite{pinson2014temporal}, MOVIE \cite{seshadrinathan2010motion}, ST-MAD \cite{vu2011spatiotemporal}, VMAF \cite{techblog} and FLOSIM \cite{manasa2016optical} and reduced-reference models like STRRED \cite{soundararajan2013video}.

No-reference (NR) VQA has also been deeply studied \cite{bovik2013automatic}. Many distortion-specific NR VQA methods \cite{yang2007perceptual, yang2005novel, kawayoke2008nr} have been designed to predict the effect of domain-relevant distortions on perceived quality. In a general model \cite{saad2014blind}, a natural scene statistics model in the DCT domain was used to train a support vector regressor to predict the effects of packet loss, MPEG-2 and H.264 compression. VIIDEO \cite{mittal2016completely} generalizes further by relying only on statistical regularities of natural videos, rather than on subjective scores or prior information about the distortion types. However, the NR VQA problem remains far from an ultimate solution.

\textit{B. Playback Interruption}

When the available bandwidth reaches a critical value (e.g. in a mobile streaming scenario), playback interruption is sometimes very difficult to avoid. Fig. \ref{fig_examples}b depicts an example of playback interruption. While the effects of rebuffering on QoE are not yet well understood, various studies have shown that the duration, frequency and location of rebuffering events severely affects QoE \cite{7025402, ghadiyaram2014study, ghadiyaram2015time, balachandran2013developing}. By making use of global rebuffering statistics, Quality of Service (QoS) models such as FTW \cite{hossfeld2011quantification} and VsQM \cite{rodriguez2012quality} have been proposed. More recent efforts \cite{ghadiyaram2015time} have sought to both model the effects of rebuffering on user QoE, and to integrate them with models of recency \cite{hands2001recency}. 

These video impairments are usually studied in isolation. For example, QoE models have either been designed for videos suffering from compression distortion or from rebuffering, but not both. This is partly due to the unavailability of suitable subjective data, along with the difficulty of combining objective video quality models and rebuffering-related information into single QoE scores. In \cite{duanmu2016sqi}, FR quality algorithms such as SSIM and MS-SSIM were combined with rebuffering information yielding the Streaming Quality Index (SQI). In \cite{singh2012quality}, the authors fed QP values and rebuffering related features into a Random Neural Network learning model to make QoE predictions. However, their method was evaluated on only 4 contents and on short video sequences of 16 seconds, did not consider longer term memory effects and did not deploy perceptually relevant VQA algorithms. This suggests the need for larger streaming-oriented subjective datasets and algorithms which collectively build on perceptually driven VQA methods, rebuffering models and other QoE-aware features. Note that HAS uses TCP, hence it is resilient to video quality degradations related to packet loss, such as glitches and other transient artifacts \cite{5404314}. As a result, the two main impairment categories that a streaming dataset should include are compression (due to the multiple encoding bitstream representations of the high-quality source content) and playback interruptions (due to throughput and buffer limitations).

We begin by describing the recently designed LIVE-Netflix Video QoE Database which contains videos suffering from temporal rate/quality changes and rebuffering events. Next, we develop Video ATLAS: a new learning framework that integrates objective VQA metrics with rebuffering-related features to conduct QoE prediction.

\section{The LIVE-Netflix rebuffering dataset}
\label{dataset}

Most existing video quality databases consider the two main video impairments (quality changes and playback interruptions) either in isolation or in an \textit{ad hoc} fashion, hampering their practical relevance. In addition, due to the difficulty of designing and carrying out large video subjective studies, many of these datasets are of quite limited size in terms of video content and/or the number of participants. We recently designed the LIVE-Netflix Video QoE Database \cite{DB_paper}, which uses a set of 8 different playout patterns on 14 diverse video contents. The video content spans a variety of content types typical of streaming applications, including action scenes, drama, cartoons and anime. We gathered approximately $5000$ subjective QoE (both continuous and retrospective) scores from 56 subjects, each participating in three 45 minute sessions.

\begin{figure}[tp]
\centerline{
\includegraphics[width = 0.8\columnwidth] {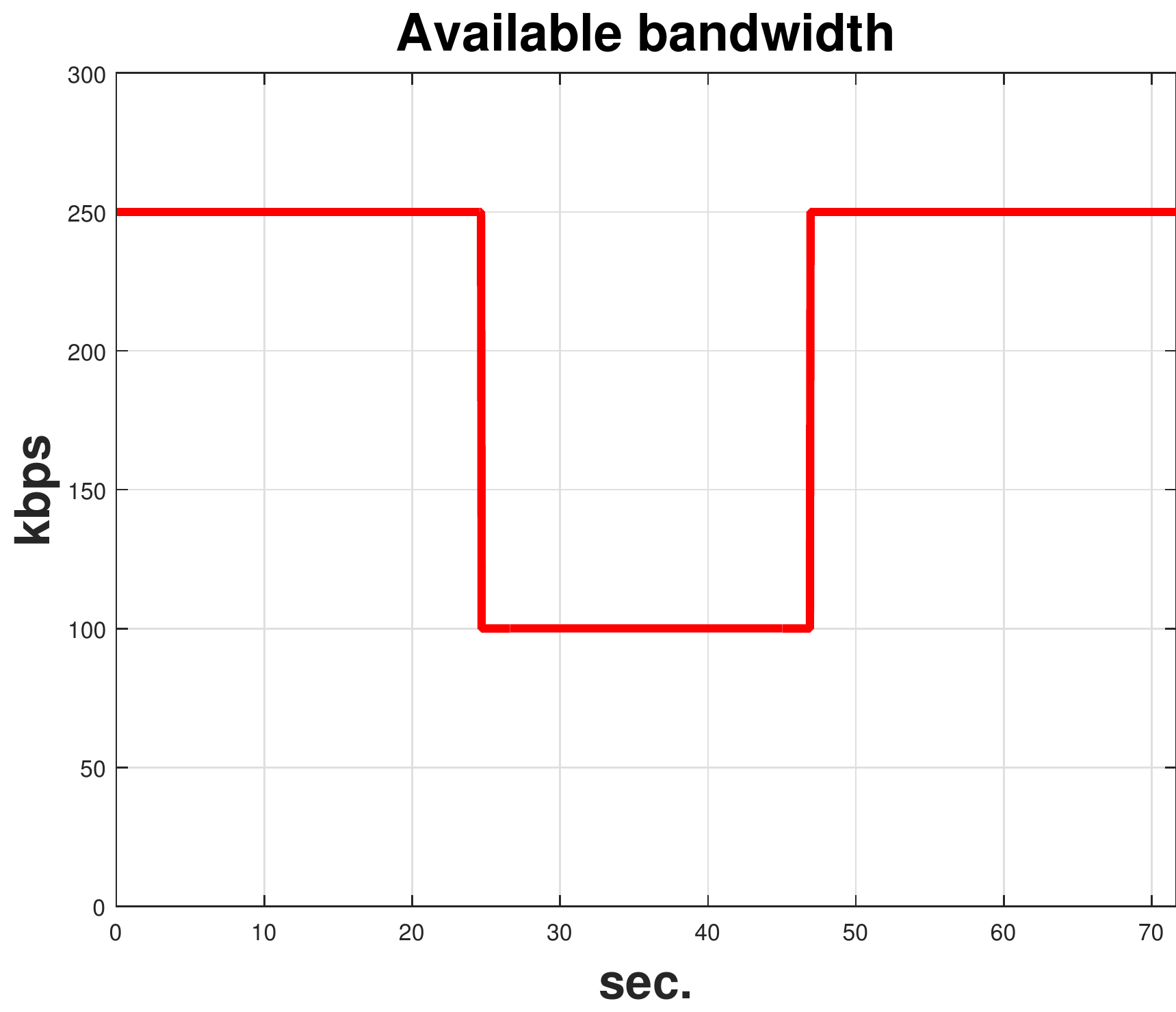}
}
\caption{Exemplar bandwidth condition in the LIVE-Netflix Video QoE Database. Horizontal axis: frame number; Vertical axis: available bandwidth in kbps.}
\label{net_cond}
\end{figure}

The playout patterns contain mixtures of static and dynamic bitrate selection strategies together with playback interruptions, assuming practical network conditions and buffer size. Figure \ref{net_cond} shows the exemplar temporal bandwidth condition. For all playout patterns, it is assumed that the available bandwidth can reach a maximum value of $250$ kbps and a minimum of $100$ kbps; a variety of playout bitrates can occur within this range. However, the buffer capacity is assumed constant over all playout patterns.
The underlying study design allows for direct comparisons between playout patterns with regards to bitrates, and the locations and the durations of playback interruptions. These playout patterns model realistic network allocation policies that content providers need to decide on. The diverse spatiotemporal characteristics and realistic playout patterns make the new LIVE-Netflix dataset a useful tool for training and evaluating video QoE predictors. This dataset consists of both public and Netflix content. The public videos together with metadata for all videos will be made available.

\section{Is objective VQA enough?}
\label{is_enough_section}

Most VQA algorithms do not consider playback interruptions. However, the increasingly pressing problem of rebuffering events in streaming applications dictates the need to quantify the effects of using (or failing to use) rebuffering-aware methods when predicting user QoE. Therefore, we selected a few important objective quality metrics and applied them on the LIVE-Netflix dataset twice. First, on the set of videos distorted only by video quality changes with normal playback ($S_{q}$). Second, on all the videos in the dataset ($S_{all}$). Then, we calculated the correlations of the prediction models against the retrospective subjective scores in the LIVE-Netflix Database to better understand the effect of including rebuffering-aware information. We used the following models: PSNR, PSNRhvs \cite{ponomarenko2007between}, SSIM \cite{wang2004image}, MS-SSIM \cite{wang2003multiscale}, NIQE \cite{mittal2013making}, VMAF \cite{techblog}, the FR version of STRRED \cite{soundararajan2013video} and GMSD \cite{xue2014gradient}. Note that for PSNRhvs \cite{ponomarenko2007between} we used the publicly available implementation of the Daala codec \cite{daala}. For the rest of the implementations, we use the publicly available implementations and all objective quality metrics were applied on the luminance component. The results are tabulated in Table \ref{isenough}.

\begin{table}[!ht]
\caption{Spearman Rank Order Correlation Coefficient (SROCC) of various image/video quality assessment (IQA/VQA) algorithms after performing mean pooling on the no rebuffering subset ($S_{q}$) and on the whole dataset ($S_{all}$).} 
\label{isenough}
\centering
\begin{tabular}{ |c|c|c|}
\hline
IQA/VQA metric & $S_{q}$ & $S_{all}$ \\
\hline
PSNR (IQA, FR) & 0.5561 & 0.5152 \\
\hline
PSNRhvs \cite{ponomarenko2007between} (IQA, FR) & 0.5841  & 0.5385 \\
\hline
SSIM \cite{wang2004image} (IQA, FR) & 0.7852 & 0.7015 \\
\hline
MS-SSIM \cite{wang2003multiscale} (IQA, FR) & 0.7532 & 0.6800 \\
\hline
NIQE \cite{mittal2013making} (IQA, NR) & 0.3960 & 0.1697  \\
\hline
VMAF \cite{techblog} (VQA, FR) & 0.7533 & 0.6097 \\
\hline
STRRED \cite{soundararajan2013video} (VQA, RR) & 0.7996 & 0.6594 \\
\hline
GMSD \cite{xue2014gradient} (IQA, FR) & 0.6476 & 0.5812 \\
\hline
\end{tabular}
\end{table}

Consider $S_{q}$, which only includes video compression artifacts. NIQE performed the worst, since it is a frame-based NR method. PSNR performed worse than all FR methods, while PSNRhvs achieved a small improvement over PSNR. The gradient-based GMSD performed worse than SSIM. STRRED yielded the best performance whereas VMAF performed poorly. Notably, STRRED performed similar to SSIM, while MS-SSIM performed worse than SSIM. This raises the following contradiction: if we consider videos that suffered only from video bitrate changes ($S_{q}$) , why would a single scale algorithm such as SSIM perform better than its multiscale counterpart and almost the same as a more sophisticated VQA model such as STRRED? We believe that when subjects are exposed to both rebuffering and quality changes, they tend to internally compare between them rather than evaluating their QoE merely based on quality changes. This makes objective video quality models less reliable, by decorrelating their performance against perceived QoE. This strongly suggests that rebuffering and bitrate changes must be considered jointly and not in isolation.

Next we consider the performance of these quality models on $S_{all}$. First, there was clearly a large drop in the performance of all models compared to $S_{q}$. Note that SSIM unexpectedly outperformed STRRED and MS-SSIM. This suggests that objective quality models are less suitable for QoE prediction on videos afflicted by interrupted playback. However, in mobile streaming applications, rebuffering events occur often. Again, this implies the need to integrate QoE-aware information into QoE prediction models. In this direction, we next describe a new learning framework which integrates objective video quality, rebuffering-related and memory features to significantly improve QoE prediction.

\section{Learning-based framework for QoE prediction}
\label{our_method}

Our proposed framework is designed to make predictions on retrospective QoE scores i.e. the subjective score given by subjects after the video playback has finished. In order to capture both video quality and to predict reactions to playback interruptions, we compute the following types of QoE-relevant input features:

1. Objective video quality scores (VQA)

During normal playback, any good video quality algorithm can be used to measure objective QoE scores. Our method allows the use of any full reference (FR) or no reference (NR) image/video quality model \cite{bovik2013automatic} as appropriate for the application context. We selected several that are both highly compute-efficient and that deliver accurate VQA predictions, rather than using compute-intensive models \cite{seshadrinathan2010motion, seshadrinathan2007structural}. Since we are focused on predicting retrospective QoE scores, a pooling strategy was chosen that collapses per-frame objective quality measurements into a single value. A number of different pooling strategies have been proposed \cite{moorthy2012video, seufert2013pool, park2013video} that capture subjective QoE aspects such as recency (whereby more recent experiences have a larger weight when making retrospective evaluations) or the peak-end effect (the worst and best parts of an event affect the QoE more). For simplicity, we deployed simple averaging of the QoE scores as suggested in \cite{seufert2013pool}, reserving recency modeling as a separate input feature.

2. Rebuffering-aware features (R$_1$ and R$_2$)

When the playback is interrupted, objective video quality algorithms are not operative. Based on previous observations regarding the effects of re-buffering \cite{pastrana2004sporadic, qi2006effect, hossfeld2011quantification, ghadiyaram2014study, ghadiyaram2015time}, we use the length of each rebuffering event measured in seconds (R$_1$) and the number of rebuffering events (R$_2$). The length of the rebuffering event(s) were normalized to the duration of each video.

3. Memory-related feature (M)

The users' QoE also depends on the recency effect. When conducting retrospective QoE prediction, we computed the time since the last rebuffering event or rate drop took place and was completed i.e. the number of seconds with normal playback at the maximum possible bitrate until the end of the video. This feature was normalized to the duration of each video.

4. Impairment duration feature (I)

While the previous features consider rebuffering and quality changes, we also computed the time (in sec.) per video over which a bitrate drop took place; following the simple notion that the relative amount of time that a video is more heavily distorted is directly related to the overall QoE. This feature was normalized to the duration of each video.

We now describe the feature extraction process. Consider all frame pairs ($i$, $i+j$), where $i$ indexes the $i$th frame of the pristine video and $i+j$ indexes the corresponding frame of the distorted video, where $j\ge 0$. If there are no rebuffering events in the distorted video then $j=0$ $\forall i$; else we determine $j$ based on the number of frozen frames up until this point for a particular video. In other words, these two frames must be synchronized in order to be able to extract meaningful objective quality measurements. Next, apply any FR IQA or VQA algorithm to measure the per-frame objective quality, then apply simple average pooling of those values, yielding a single quality-predictive feature that will be used later. In addition, all the other features are collected, assuming that for retrospective QoE prediction, the number of rebuffered frames as well as the locations of the bitrate changes are known. Note that for some VQA methods, adjacent frames may be needed to compute frame differences. In that case we ensure that all frame differencing takes place between two consecutive frames that both have normal playback. If an NR method is used, it is computed only on unstalled frame(s).

After collecting all the features computed on each video, we then deployed a learning-based approach where the subjective data and the input features were used to train a regression engine. Note that no constraint was placed on which objective quality algorithm or regression model is used. In our experiments, we studied the performance of our proposed approach across different regression and IQA/VQA models. The final output of our overall system is a single retrospective QoE score on each input test video. 

\section{Training and evaluation of the proposed framework}
\label{our_results}

\subsection{Experiments on the LIVE-Netflix Video QoE Database}

To evaluate the proposed method on the LIVE-Netflix Video QoE Database, we conducted two different experiments. The first one (Experiment 1) consisted of creating two disjoint content sets: one for training and one for testing. Within each content (training or testing), all patterns were used for training or testing. While this is a common approach used to account for content dependencies in learning-based VQA methods, it may also occur that the different ``distortions" or playout patterns induce pattern dependencies, resulting in overestimation of the true predictive power of a learning-based method. To examine pattern independence we also conducted a second experiment (Experiment 2), where we picked one of the playout patterns as a test pattern and the rest as training patterns. Thus, for each testing pattern there were 14 test points (one for each content) and 98 testing points. On both tests, we applied a regression model (e.g. Random Forest regression) to predict the QoE scores of the test set given the input training features and MOS scores. We excluded the subjective scores gathered from the three training videos. Since our model does not produce continuous scores, we used only the retrospective QoE scores from all 14 test contents.

To demonstrate the behavior of Video ATLAS we evaluated it using several different types of regression models \cite{sklearn}: linear models (Ridge and Lasso regression), Support Vector Regression (SVR) using a rbf kernel and ensemble methods such as Random Forest (RF), Gradient Boosting (GB) and Extra Trees (ET) regression. For the ensemble methods, feature normalization was not required, but we preprocessed the features for all regression models by mean subtraction and scaling to unit variance. Note that we computed the data mean and variance in the feature transformation step using only the training data. For each of the regression models, we determined the best parameters using 10-fold cross validation on the training set. This process was repeated on all possible train/test splits.

After each of the regression models was trained, we applied regression on the test features to make QoE predictions. Then, we correlated the regressed values with the MOS scores in the test set and calculated the Spearman Rank Order Correlation Coefficients (SROCC) and the Pearson Linear Correlation Coefficients (LCC). The former measures the monotonicity of the regressed values and the latter the linearity of the output, which is highly desirable since it describes the degree of simplicity of a trained model. Before computing the LCC, we first applied a non-linear regression step on the output QoE scores of our method, as suggested in \cite{bt500}.

\subsubsection{Experiment 1: Testing for Content Independence}

We conducted 1000 different trials, each using a random $80\%$ train and $20\%$ test split of the video content. To avoid content dependencies, we select $80\%$ of the 14 contents in the database as the training contents (11 training contents) and the rest as the testing contents (3 testing contents). For direct comparison, we used a pre-generated set of train/test indices. The SROCC and LCC calculations were repeated on each of the trials yielding a distribution of SROCC and LCC values for all possible train/test content combinations. Taking the median value of this distribution of correlation scores yields a single number describing the performance level of the proposed method. Table \ref{results_SROCC_LCC} shows the SROCC and LCC results after $1000$ trials. 
\begin{table*}[!ht]
\centering
\caption{Results on different image/video quality assessment algorithms (IQA/VQA) after performing mean pooling on the objective quality metrics. Top: Spearman Rank Order Correlation Coefficient (SROCC); Bottom: Pearson's linear correlation coefficient (LCC). For each metric we report the median SROCC/LCC before regression (BR) using only the IQA/VQA metric and the SROCC/LCC values after regression when using different objective quality metrics and regression models. The last column contains the average of the SROCC/LCC values across all quality metrics for each regression model. All results are reported on $1000$ pre-generated $80\%$ train and $20\%$ test splits. Best regression model per quality metric is denoted by bold; best result overall denoted by italic and bold.} 
\label{results_SROCC_LCC}
    \scalebox{0.9}
    {
    \begin{tabular}{| c | c | c | c | c | c | c | c | c | c | c | c | c | c | c | c | c |}
    \hline
    VQA & PSNR & PSNRhvs \cite{ponomarenko2007between} & SSIM \cite{wang2004image} & MS-SSIM \cite{wang2003multiscale} & NIQE \cite{mittal2013making} & VMAF \cite{techblog} & STRRED \cite{soundararajan2013video} & GMSD \cite{xue2014gradient} & mean \\ \hline
    BR & 0.6074 & 0.6252 & 0.6748 & 0.6557 & 0.1391 & 0.6043 & 0.6348 & 0.6496 & 0.5734 \\ \hline
    Ridge & \textbf{0.6687} & \textbf{0.6817} & 0.7565 & 0.7461 & 0.4130 & 0.6278 & 0.7957 & \textbf{0.6948} & 0.6730 \\ \hline
    Lasso & 0.6496 & 0.6687 & 0.7461 & 0.7383 & 0.4191 & \textbf{0.6409} & 0.7983 & 0.6922 & 0.6691 \\ \hline
    SVR & 0.6313 & 0.6417 & 0.8252 & 0.8226 & 0.6730 & 0.6026 & \textbf{0.8704} & 0.6878 & \textbf{0.7193} \\ \hline
    ET & 0.4265 & 0.4387 & \textbf{0.8547} & \textbf{\textit{0.8752}} & \textbf{0.7530} & 0.4756 & 0.8439 & 0.4527 & 0.6400 \\ \hline
    RF & 0.4931 & 0.5312 & 0.8088 & 0.8154 & 0.6222 & 0.4930 & 0.8104 & 0.5417 & 0.6395 \\ \hline
    GB & 0.4830 & 0.4944 & 0.7990 & 0.7899 & 0.5878 & 0.5145 & 0.8032 & 0.5000 & 0.6215 \\ \hline
    \end{tabular}
    }\vspace{2mm}
        \scalebox{0.9}
    {
    \begin{tabular}{| c | c | c | c | c | c | c | c | c | c | c | c | c | c | c | c | c |}
    \hline
    VQA & PSNR & PSNRhvs \cite{ponomarenko2007between} & SSIM \cite{wang2004image} & MS-SSIM \cite{wang2003multiscale} & NIQE \cite{mittal2013making} & VMAF \cite{techblog} & STRRED \cite{soundararajan2013video} & GMSD \cite{xue2014gradient} & mean \\ \hline
    BR & 0.6048 & 0.6534 & 0.7288 & 0.7104 & 0.3752 & 0.7561 & 0.7213 & 0.6861 & 0.6545 \\ \hline
    Ridge & 0.8145 & 0.8224 & 0.8531 & 0.8517 & 0.5984 & 0.8158 & 0.8703 & 0.8254 & 0.8064 \\ \hline
    Lasso & \textbf{0.8192} & \textbf{0.8312} & 0.8558 & 0.8514 & 0.6034 & \textbf{0.8292} & 0.8719 & \textbf{0.8374} & 0.8124 \\ \hline
    SVR & 0.7939 & 0.8016 & 0.9073 & 0.8973 & 0.7633 & 0.7742 & \textbf{\textit{0.9358}} & 0.8106 & \textbf{0.8355} \\ \hline
    ET & 0.6325 & 0.6392 & \textbf{0.9186} & \textbf{0.9289} & \textbf{0.8407} & 0.6808 & 0.9088 & 0.6869 & 0.7796 \\ \hline
    RF & 0.6767 & 0.6922 & 0.8905 & 0.8868 & 0.7182 & 0.6591 & 0.8770 & 0.7026 & 0.7629 \\ \hline
    GB & 0.6744 & 0.7060 & 0.8661 & 0.8546 & 0.7143 & 0.7115 & 0.8678 & 0.7043 & 0.7624 \\ \hline
    \end{tabular}
    }
\end{table*}

First, note that both the SROCC and the LCC were improved when using the regression scheme for all quality metrics and for at least one regression model type. For VMAF, PSNR, PSNRhvs and GMSD the regression result did not improve using every regressor. However, the improvements of SSIM, MS-SSIM, NIQE and STRRED were remarkably higher for all the regression models. MS-SSIM using ET yielded the best overall performance in terms of SROCC, while STRRED using SVR yielded the best LCC value. STRRED is an information-theoretic approach to VQA that builds on the innovations in \cite{sheikh2006image, sheikh2005visual}. It achieves quality prediction efficiency without the need to compute motion vectors, unlike \cite{seshadrinathan2007structural, seshadrinathan2010motion}. Regarding improvements in terms of LCC, all regression models improved most of the quality metrics. These observations support the argument that introducing an effective regression scheme into the QoE process has a large positive impact on QoE prediction over a wide range of leading video quality models.

\begin{figure}[tp]
\centerline{
\includegraphics[width=0.5\columnwidth] {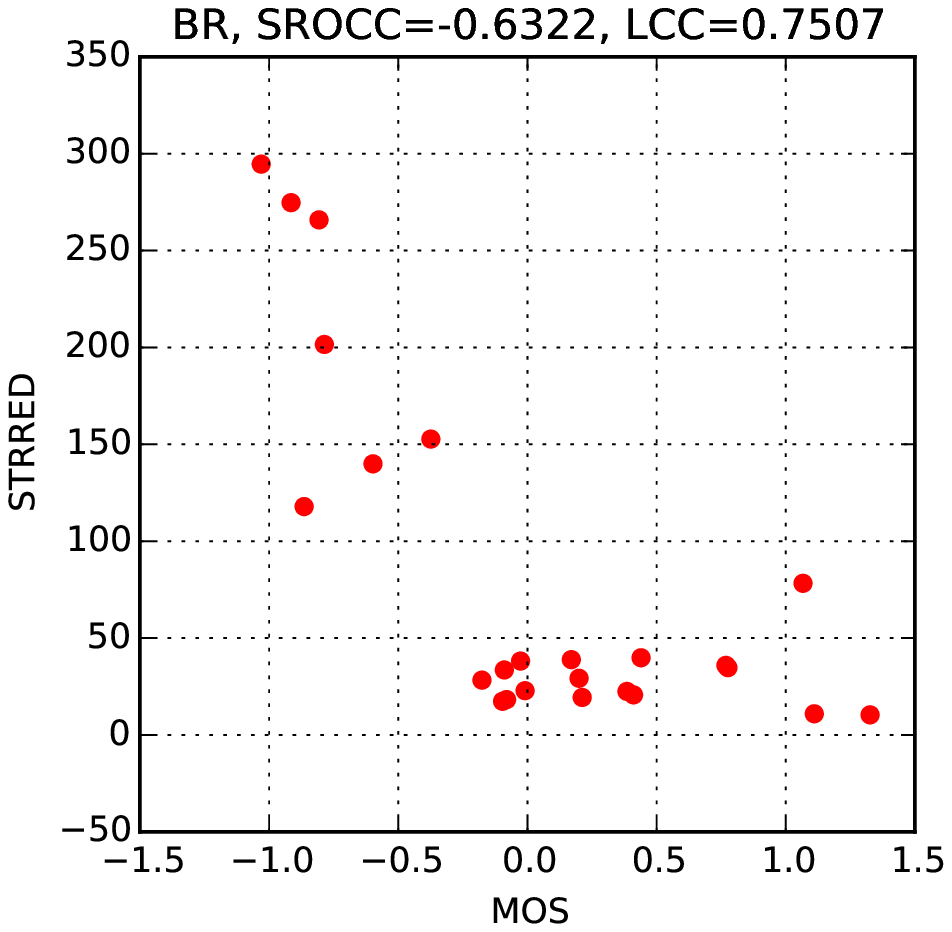}
\includegraphics[width=0.5\columnwidth] {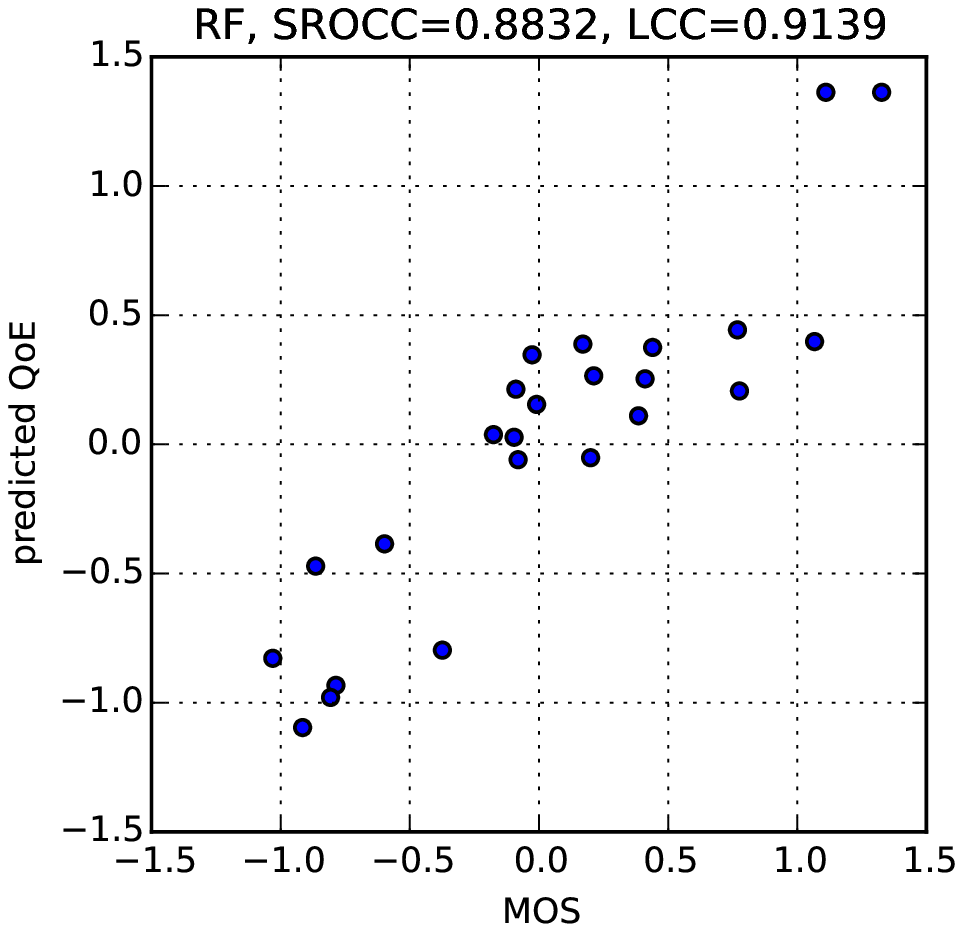}
}
\caption{MOS scores (horizontal axis) against predicted QoE scores (vertical axis) on one test set when using STRRED and Random Forest regression. Left: without regression when using only STRRED to predict the QoE; right: QoE scores after regression when using all features. When using the regressed values the monotonicity may change sign (here it becomes increasing) and the scale of the horizontal axis may also change.}
\label{visual_example}
\end{figure}

To demonstrate the overall improvements delivered by the learned regression models, we also calculated the average SROCC and LCC values for the BR case and for each regression model separately (see the last columns of Table \ref{results_SROCC_LCC}). In both cases, the SVR regressor achieved the highest average performance followed by Ridge. The performance of the Ridge and Lasso models was somewhat higher than that of the RF and ET, while GB yielded the worst performance across all regression models, although it was still higher than the average performance of BR, which was notably low in the case of NIQE. Next, we visually demonstrate the effect of the proposed learning framework (in Fig. \ref{visual_example}) for the case of STRRED and the Random Forest regression model. Clearly, the predicted QoE significantly improved both in terms of monotonicity and linearity.

\begin{figure}[tp]
\centerline{
\includegraphics[width=0.5\columnwidth, height=4.3cm] {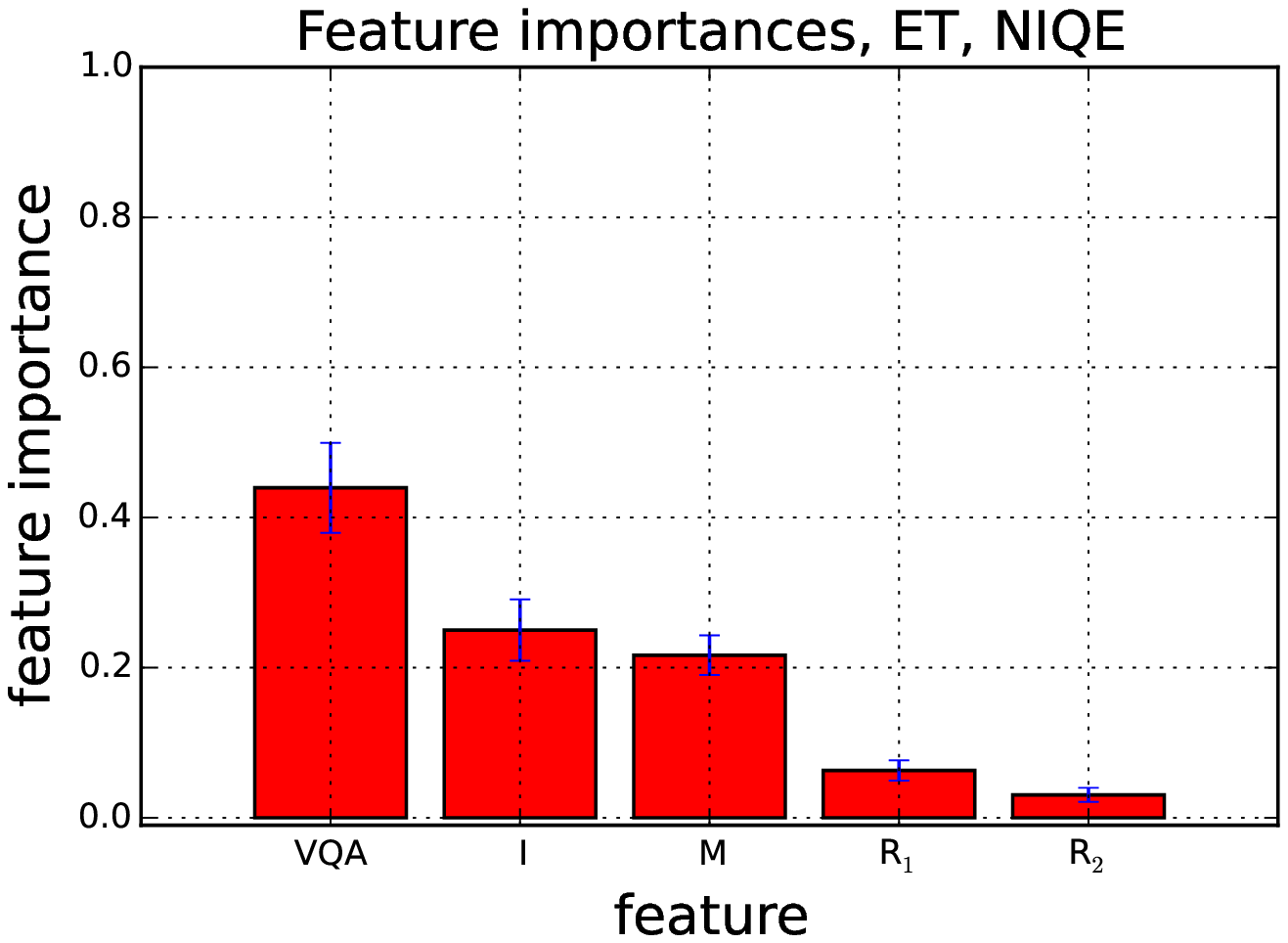}
\includegraphics[width=0.5\columnwidth,  height=4.3cm] {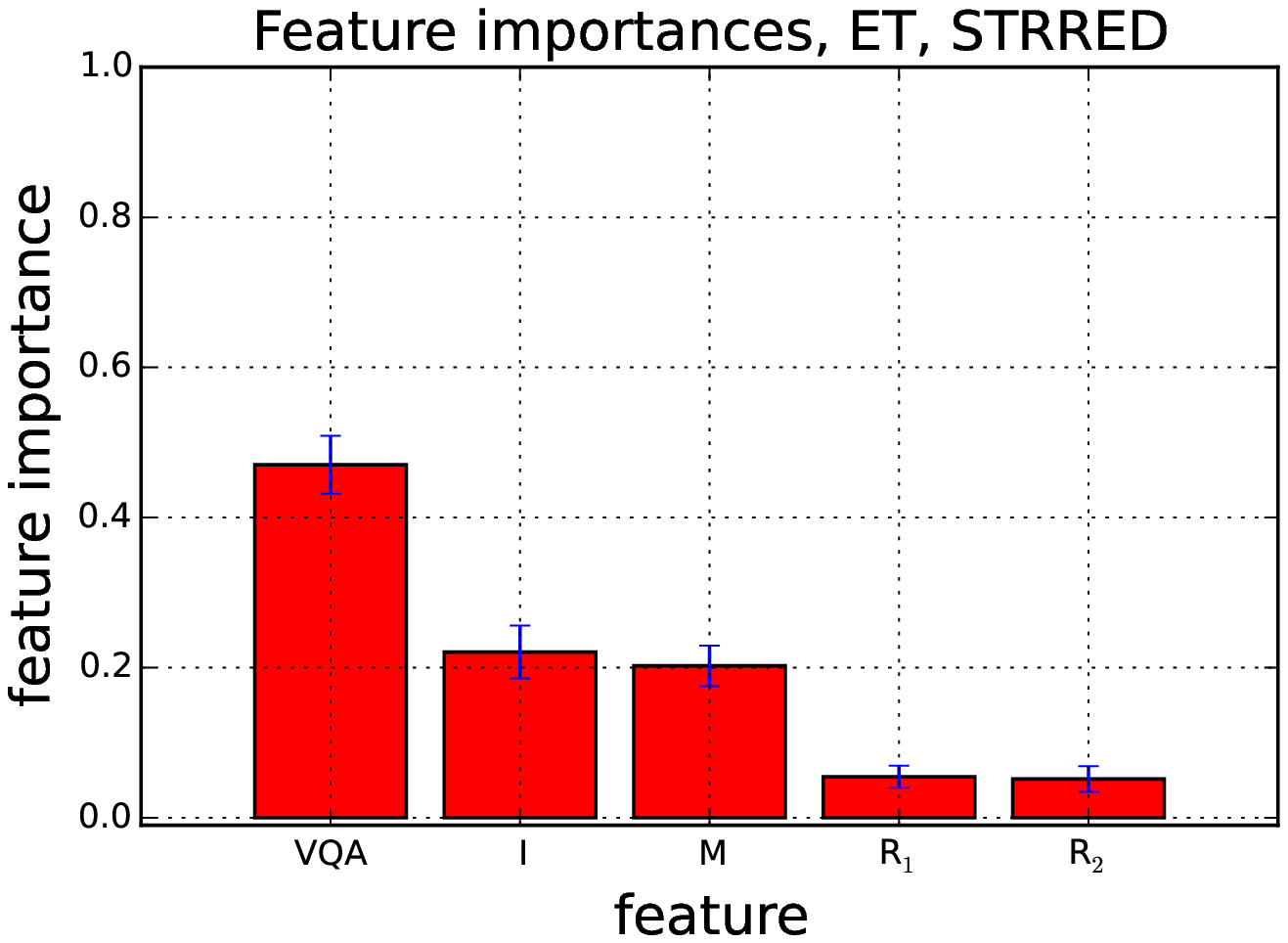}
}
\caption{Feature importances using NIQE (left) and STRRED (right) after 1000 random train/test splits using the best regressor (ET). Horizontal axis: feature labels; vertical axis: feature importance normalized to 1.}
\label{feat_imps}
\end{figure}

While our proposed system deploys features that collectively deliver excellent results, it is interesting to analyze the relative feature contributions. One way to study the feature importances is by a tree-based method, as follows. First, we picked the best and the worst performing quality models before regression (when evaluated on the whole database), i.e., STRRED and NIQE, along with the highest performing SVR regression model (in terms of SROCC). Figure \ref{feat_imps} shows the feature importances after $1000$ pre-generated train/test splits.
Clearly, the video quality model used plays an important role in QoE prediction. The memory feature also has a strong contribution since for retrospective QoE evaluation, recent experiences are a strong QoE indicator. The rebuffering features delivered an important but somewhat smaller contribution. For retrospective QoE evaluations and distinct impairment events such as rebuffering, the lower contribution of the R$_1$ feature (rebuffering duration) may possibly be explained by the duration neglect effect \cite{hands2001recency}: subjects may remember that a rebuffering event occurred, but may not be sensitive to its duration. However, as demonstrated earlier, both tested video quality models were greatly improved in terms of both SROCC and LCC when combined with Video ATLAS. Since NIQE is not a very good video quality predictor (although it is a very effective still picture quality predictor), the importance of the VQA feature was lower while the importance of the I and M features was relatively higher as compared to STRRED.

To further investigate the effects of those feature types on the retrospective QoE prediction task, we experimented further by using different feature subsets, and recording the QoE prediction performance of each. First, consider the following feature subsets:
\begin{enumerate}
\item{individual feature subsets: VQA(1), M(2), \\ I(3) and R$_1$+R$_2$(4)}
\item{2 feature types subsets: VQA+M(5) and VQA+I(6)}
\item{$\ge3$ subsets: VQA+M+R$_2$(7), M+R$_1$+R$_2$(8), M+I+R$_1$\\+R$_2$(9), VQA+I+R$_1$+R$_2$(10), VQA+M+R$_1$+R$_2$(11) and VQA+M+I+R$_1$+R$_2$(12)}
\end{enumerate}
The SROCC and LCC results are shown in Table \ref{results_feat_subset}, where we selected STRRED as the quality prediction model. Clearly, when using the individual components as features, the QoE prediction result was maximized when using VQA but was still very low, especially for other components such as M. Notably, the regression performance for the VQA subset was maximized in the case of the Ridge and Lasso linear regressions, but for the M (memory) and R$_1$+R$_2$ (rebuffering) feature types, the performance was greatly reduced using those regression models compared to SVR, ET, RF and GB. This may be explained by the fact that the design of IQA/VQA algorithms such as STRRED ultimately aims for linear/explainable models. By contrast, the memory or rebuffering-aware features are highly non-linear, hence non-linear regression models may be expected to perform better.

We now move on to the different feature combinations and their effect on QoE prediction. First, note that when VQA is removed from the feature set (e.g. in columns 8 and 9) the prediction performance dropped considerably. Meanwhile, using only two features (VQA and M in column 5) we were able to achieve better prediction results than with any other combination of 2 feature types (or a single feature). This again strongly supports the importance of memory/recency effects on QoE when viewing longer video sequences. Regarding the regression models, Ridge and Lasso gave very similar performances when using fewer feature types, but as the number of features grew, Lasso yielded better results. Overall, the combination of all feature types gave the best performance over most regression models. This suggests that a successful QoE prediction model should consider diverse QoE-aware features in order to better approximate subjective QoE.

\begin{table*}[!ht]
\centering
\caption{Results on different feature subsets when STRRED was used as the quality metric (VQA) and mean pooling was applied. Top: SROCC; Bottom: LCC. All results are reported over $1000$ pre-generated $80\%$ train and $20\%$ test splits. The best regression model per feature subset is denoted by bold; best result overall (SVR) denoted by italic and bold. The feature subsets are indexed as follows: VQA(1), M(2), I(3), R$_1$+R$_2$(4), VQA+M(5), VQA+I(6), VQA+M+R$_2$(7), M+R$_1$+R$_2$(8), M+I+R$_1$+R$_2$(9), VQA+I+R$_1$+R$_2$(10), VQA+M+R$_1$+R$_2$(11) and VQA+M+I+R$_1$+R$_2$(12).}
\label{results_feat_subset}
    \scalebox{0.96}
    {
    \begin{tabular}{| c | c | c | c | c | c | c | c | c | c | c | c | c | c | c | c | c | c | c | c |}
    \hline
    Features & 1 & 2 & 3 & 4 & 5 & 6 & 7 & 8 & 9 & 10 & 11 & 12 \\ \hline
    Ridge & \textbf{0.6348} & 0.2296 & 0.2700 & 0.3094 & 0.6000 & 0.6235 & 0.7870 & 0.4105 & 0.4172 & 0.7878 & 0.7735 & 0.7957 \\ \hline
    Lasso & \textbf{0.6348} & 0.2296 & 0.2700 & 0.3243 & 0.6304 & \textbf{0.6417} & 0.7991 & 0.4075 & 0.3955 & 0.8013 & 0.7991 & 0.7983 \\ \hline
    SVR & 0.5748 & 0.3807 & \textbf{0.2758} & \textbf{0.3740} & 0.7322 & 0.5878 & \textbf{0.8183} & 0.4210 & 0.4839 & \textbf{0.8543} & \textbf{0.8122} & \textit{\textbf{0.8704}} \\ \hline
    ET & 0.5074 & 0.3076 & 0.2345 & 0.2993 & 0.7431 & 0.5962 & 0.7496 & 0.3119 & 0.3924 & 0.8348 & 0.7574 & 0.8435 \\ \hline
    RF & 0.5304 & \textbf{0.3961} & 0.2713 & 0.3218 & \textbf{0.7537} & 0.5691 & 0.7633 & 0.4126 & 0.4656 & 0.8074 & 0.7708 & 0.8096 \\ \hline
    GB & 0.5691 & 0.3905 & 0.2658 & 0.3527 & 0.7461 & 0.6001 & 0.7668 & \textbf{0.4355} & \textbf{0.4984} & 0.8070 & 0.7607 & 0.8036 \\ \hline
    \end{tabular}
    }\vspace{2mm}
    \scalebox{0.96}
    {    
        \begin{tabular}{| c | c | c | c | c | c | c | c | c | c | c | c | c | c | c | c | c | c | c |}
    \hline
    Features & 1 & 2 & 3 & 4 & 5 & 6 & 7 & 8 & 9 & 10 & 11 & 12 \\ \hline
    Ridge & \textbf{0.7213} & 0.4507 & 0.3049 & 0.2930 & 0.7141 & 0.6475 & 0.7610 & 0.4602 & 0.6247 & 0.7854 & 0.7590 & 0.8703 \\ \hline
    Lasso & \textbf{0.7213} & 0.4507 & 0.3049 & 0.2956 & 0.7348 & \textbf{0.6956} & 0.7870 & 0.4592 & 0.6201 & 0.8055 & 0.7868 & 0.8719 \\ \hline
    SVR & 0.6454 & 0.4325 & 0.3148 & 0.3169 & \textbf{0.8133} & 0.6472 & \textbf{0.8510} & 0.4497 & 0.6959 & 0.8945 & \textbf{0.8392} & \textit{\textbf{0.9358}} \\ \hline
    ET & 0.5407 & 0.3754 & 0.3110 & 0.3138 & 0.7620 & 0.6031 & 0.7596 & 0.3899 & 0.6173 & \textbf{0.9004} & 0.7659 & 0.9090  \\ \hline
    RF & 0.5685 & 0.4451 & \textbf{0.3528} & \textbf{0.3261} & 0.7794 & 0.6024 & 0.7862 & \textbf{0.4706} & 0.6966 & 0.8686 & 0.7975 & 0.8742 \\ \hline
    GB & 0.6287 & \textbf{0.4514} & 0.3514 & 0.3141 & 0.7755 & 0.6269 & 0.7904 & 0.4751 & \textbf{0.7413} & 0.8665 & 0.7865 & 0.8686 \\ \hline
    \end{tabular}
    } 
\end{table*}

\begin{table*}[!ht]
\caption{Results when using various image/video quality assessment algorithms (IQA/VQA) after performing mean, hysteresis and VQ pooling. Left: SROCC; Right: LCC. We used Video ATLAS and all 5 input features. All results are reported over $1000$ pre-generated $80\%$ train and $20\%$ test splits. For each quality model, we selected only the best performing regression model (in terms of SROCC). The best pooling method for each video quality model is denoted by bold and the best overall performance is denoted by bold and italic.} 
\label{poolus_srocc_lcc}
\centering
\begin{tabular}{ |c|c|c|c|}
\hline
IQA/VQA metric & mean & hysteresis & VQ \\
\hline
PSNR & 0.6687 & 0.6687 & \textbf{0.6817} \\
\hline
PSNRhvs \cite{ponomarenko2007between} & 0.6817 & 0.6878 & \textbf{0.6965} \\
\hline
SSIM \cite{wang2004image} & \textbf{0.8547} & 0.8470 & 0.7887 \\
\hline
MS-SSIM \cite{wang2003multiscale} & \textbf{0.8752} & 0.8722 & 0.7743 \\
\hline
NIQE \cite{mittal2013making} & \textbf{0.7530} & 0.7513 & 0.6591 \\
\hline
VMAF \cite{techblog} & \textbf{0.6409} & 0.6226 & 0.6400 \\
\hline
STRRED \cite{soundararajan2013video} & 0.8704 & \textbf{\textit{0.8800}} & 0.8687 \\
\hline
GMSD \cite{xue2014gradient} & \textbf{0.6948} & 0.6800 & 0.6843 \\
\hline
\end{tabular}
\begin{tabular}{ |c|c|c|c|}
\hline
IQA/VQA metric & mean & hysteresis & VQ \\
\hline
PSNR & 0.8145 & 0.8173 & \textbf{0.8231} \\
\hline
PSNRhvs \cite{ponomarenko2007between} & 0.8224 & 0.8254 & \textbf{0.8635} \\
\hline
SSIM \cite{wang2004image} & \textbf{0.9186} & 0.9121 & 0.8804 \\
\hline
MS-SSIM \cite{wang2003multiscale} & \textbf{0.9289} & 0.9281 & 0.8700 \\
\hline
NIQE \cite{mittal2013making} & 0.8407 & \textbf{0.8495} & 0.7678 \\
\hline
VMAF \cite{techblog} & \textbf{0.8292} & 0.8136 & 0.8202 \\
\hline
STRRED \cite{soundararajan2013video} & 0.9358 & \textbf{\textit{0.9390}} & 0.9317 \\
\hline
GMSD \cite{xue2014gradient} & \textbf{0.8254} & 0.8062 & 0.8191 \\
\hline
\end{tabular}
\end{table*}

\begin{figure*}[tp]
\centerline{
\includegraphics[width=\columnwidth] {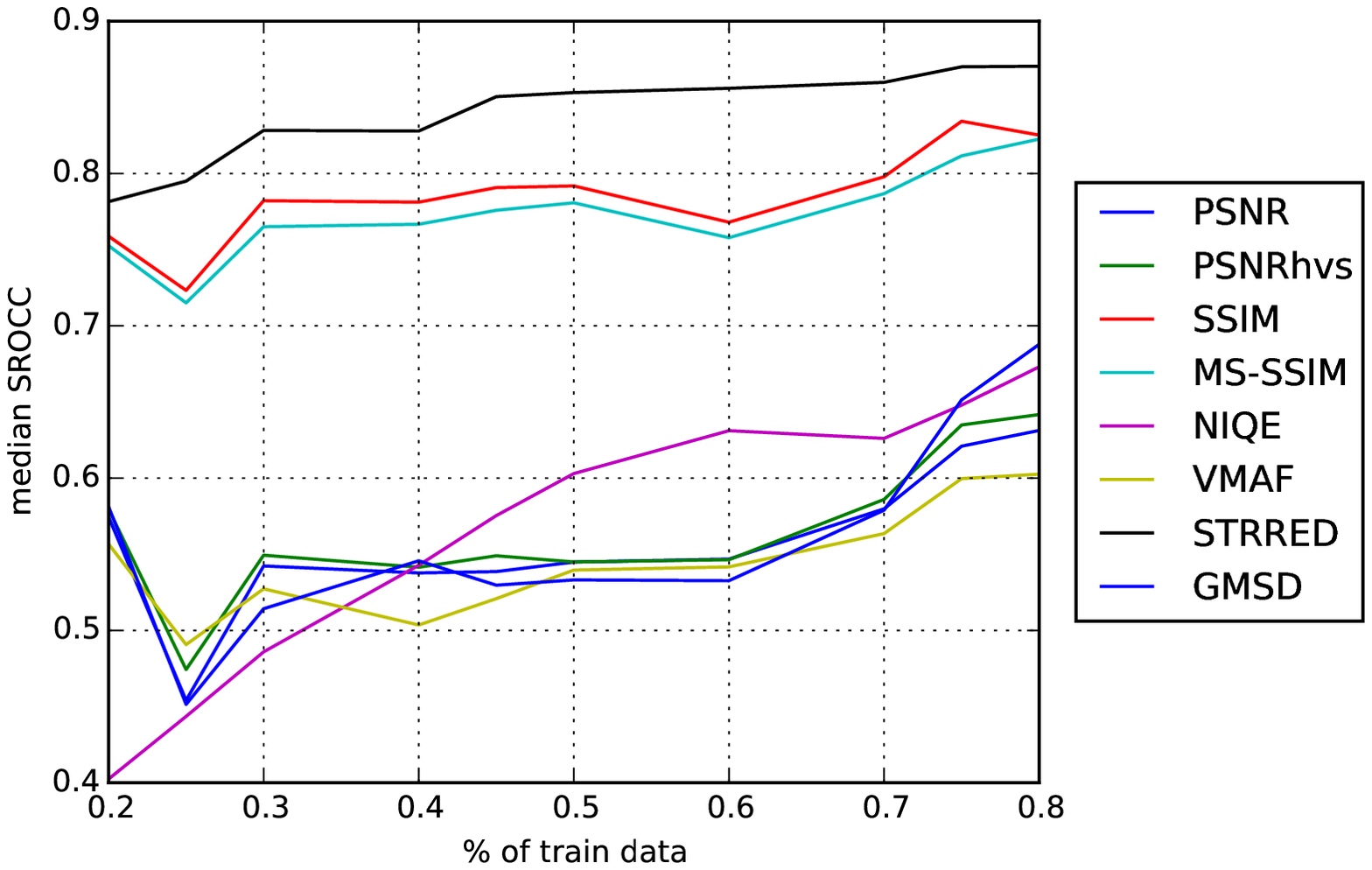}
\includegraphics[width=\columnwidth] {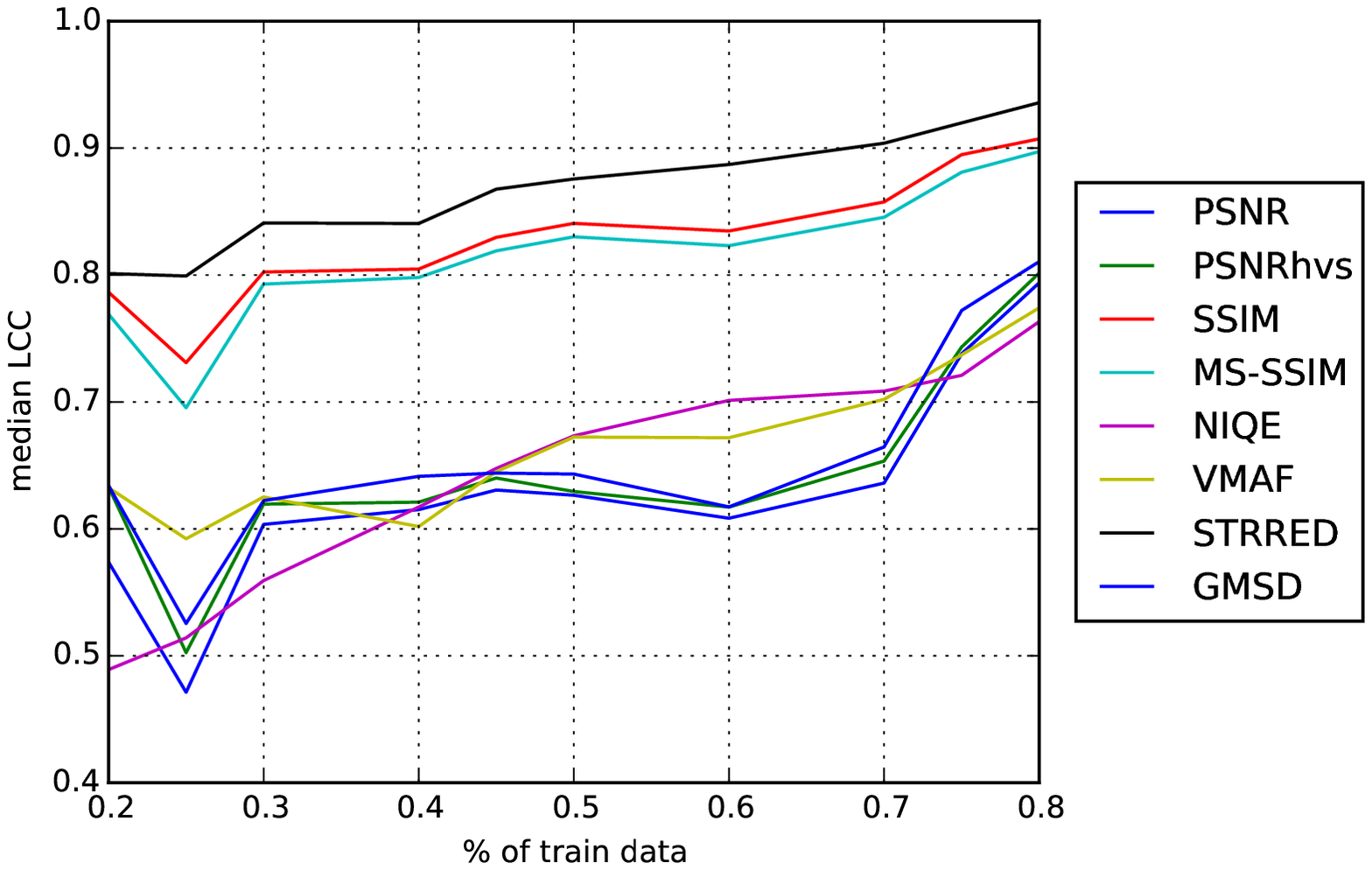}
}
\caption{Prediction monotonicity (median SROCC) and performance (median LCC) after 1000 random train/test splits as the amount of training data was varied for different objective video quality models. The ET regression model was used.}
\label{across_train}
\end{figure*}

We now analyze the effects of the amount of training data used in the regression scheme on QoE prediction. By varying the percent of training data in the train/test split, we repeated the same process as before, over $1000$ random trials. Figure \ref{across_train} shows how the SROCC changed when the amount of training data varied between $0.2$ (2 training contents) and $0.8$ (11 training contents). Clearly, the prediction performance increased when the available training data was increased. The best performance in terms of SROCC and LCC was reached when MS-SSIM was used as the quality model. Note that while NIQE performed the worst before applying regression, now it performed better than VMAF, GMSD, PSNR and PSNRhvs when the ratio of the train/test split was larger than 0.4. Notably, the SROCC performance of GMSD, PSNR and PSNRhvs did not significantly vary until the train/test split was larger than 0.6. By contrast, STRRED, SSIM and MS-SSIM delivered good results (in terms of SROCC and LCC) when only a small amount of training data was used.

We also experimented with the type of pooling that is applied on the quality metric before it is used in the regression framework. We combined all features and used the pre-generated $80\%$ train and $20\%$ test splits. To collapse the frame-based objective quality scores to a single summary VQA score, we applied the hysteresis pooling method in \cite{seshadrinathan2011temporal} and the VQ pooling method in \cite{park2013video}. The former combines past and future quality scores within a window, while the latter clusters the video frames into low and high quality regions and weights their contributions to the overall VQA score. The results are tabulated in Table \ref{poolus_srocc_lcc}. For the mean pooling case, we used the results reported in Table \ref{results_SROCC_LCC}.

Given the results in Table \ref{poolus_srocc_lcc}, we observed that the use of temporal pooling strategies other than mean pooling did not always improve QoE prediction \cite{seufert2013pool}. Of the 8 metrics we reported, only 3 were improved in terms of SROCC and 4 in terms of LCC. Further, these improvements were not very significant, with the exception of PSNR and PSNRhvs.

\begin{figure*}[tp]
\centerline{
\includegraphics[width=0.5\columnwidth] {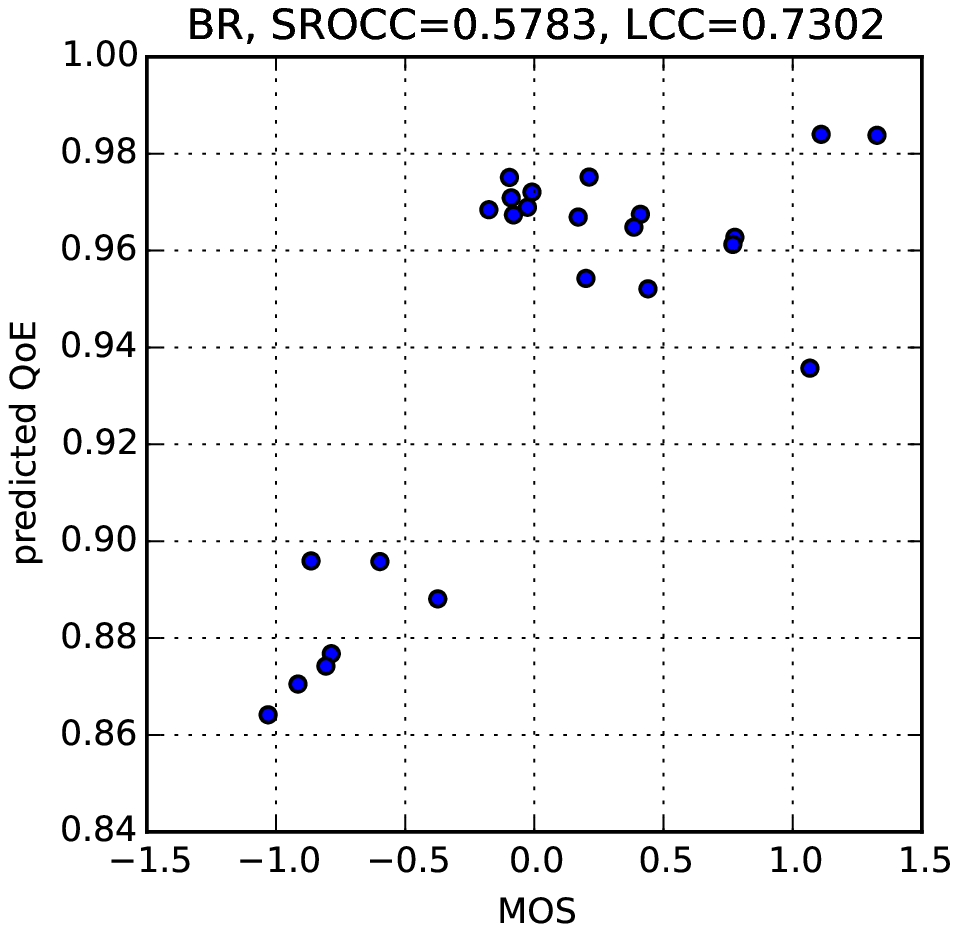}
\includegraphics[width=0.5\columnwidth] {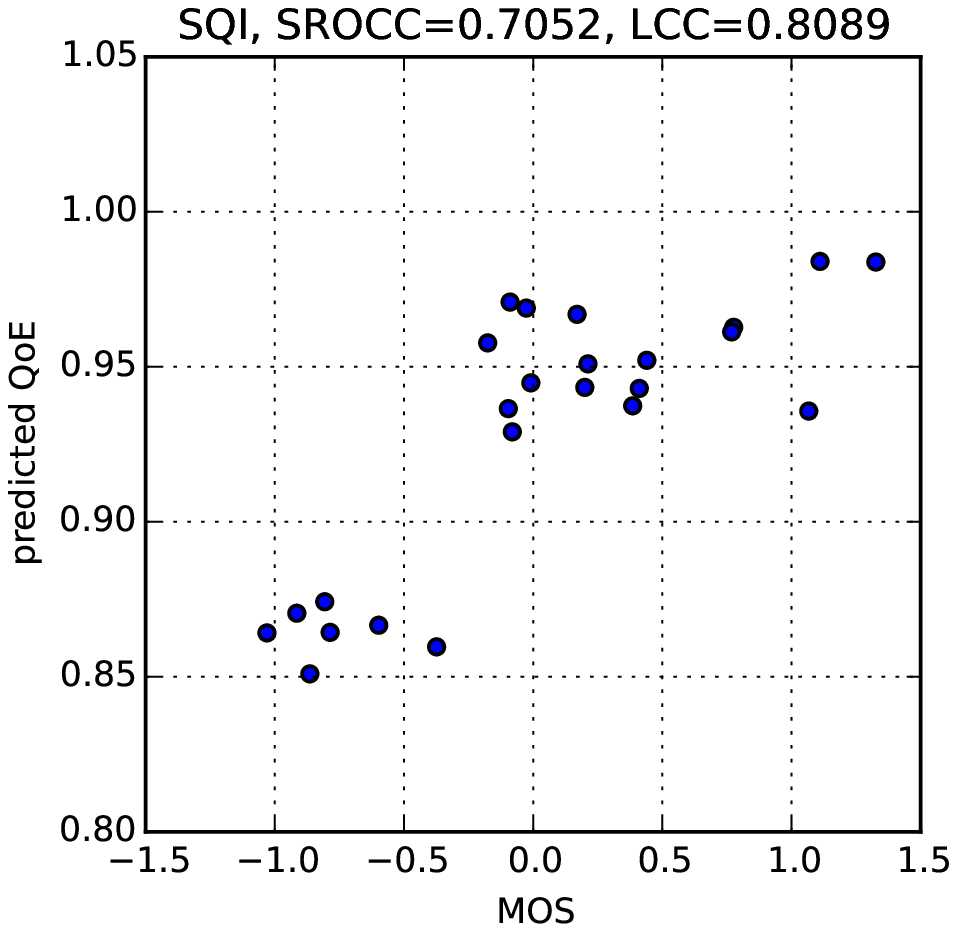}
\includegraphics[width=0.5\columnwidth] {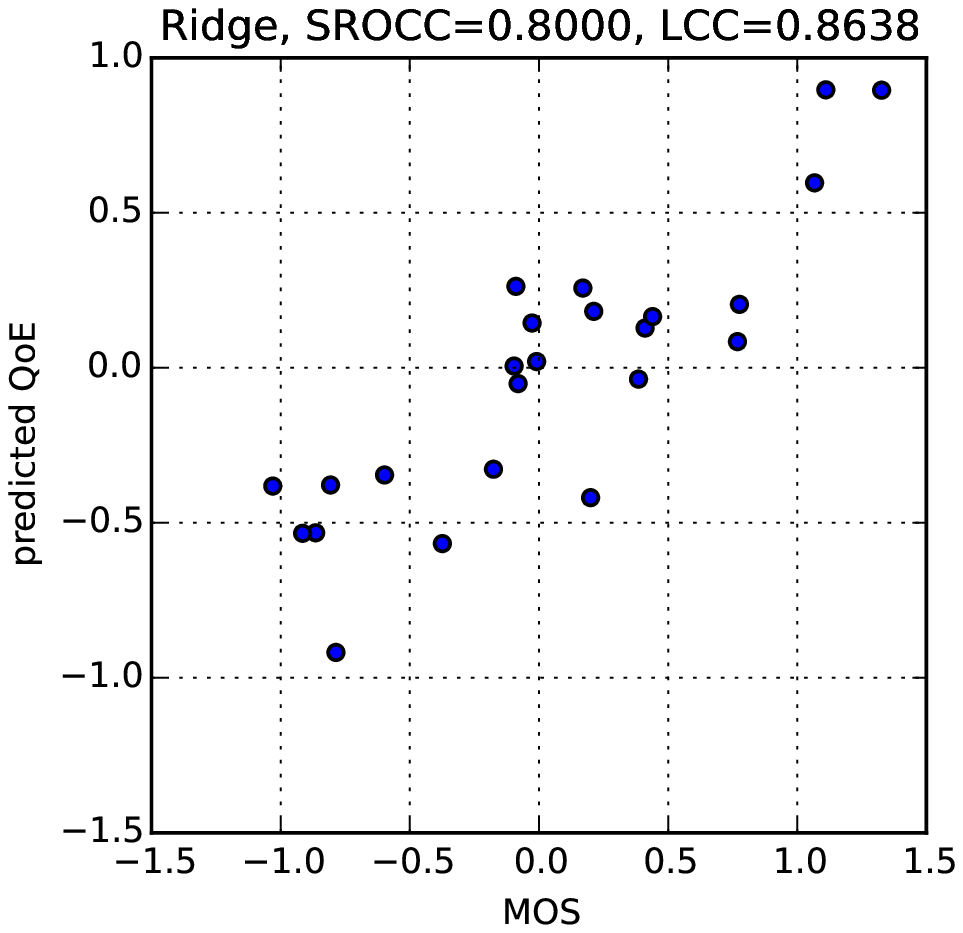}
\includegraphics[width=0.5\columnwidth] {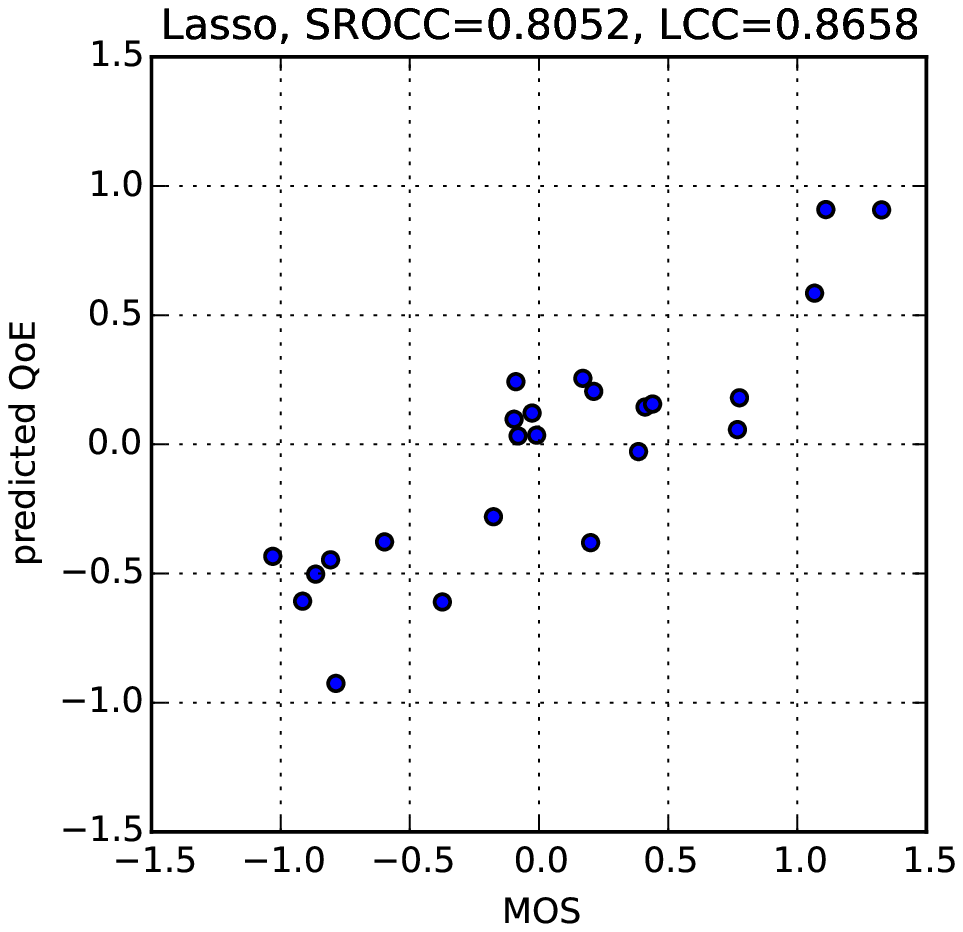}
}
\centerline{
\includegraphics[width=0.5\columnwidth] {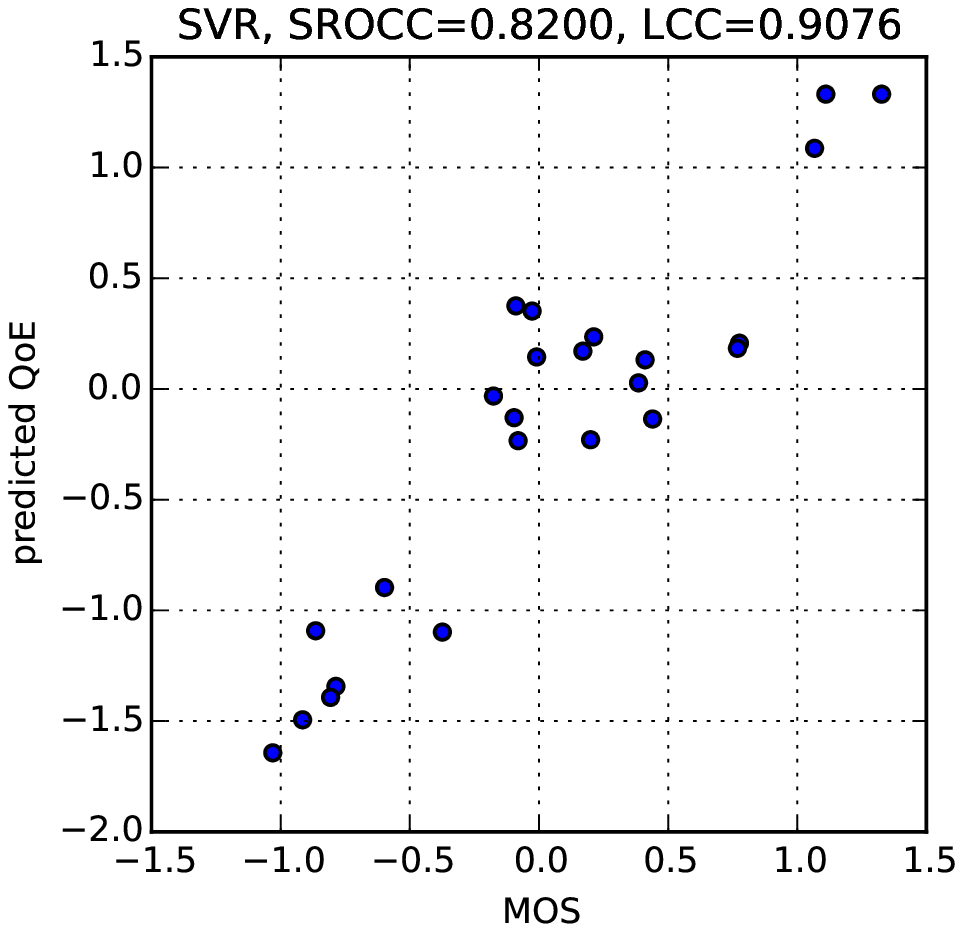}
\includegraphics[width=0.5\columnwidth] {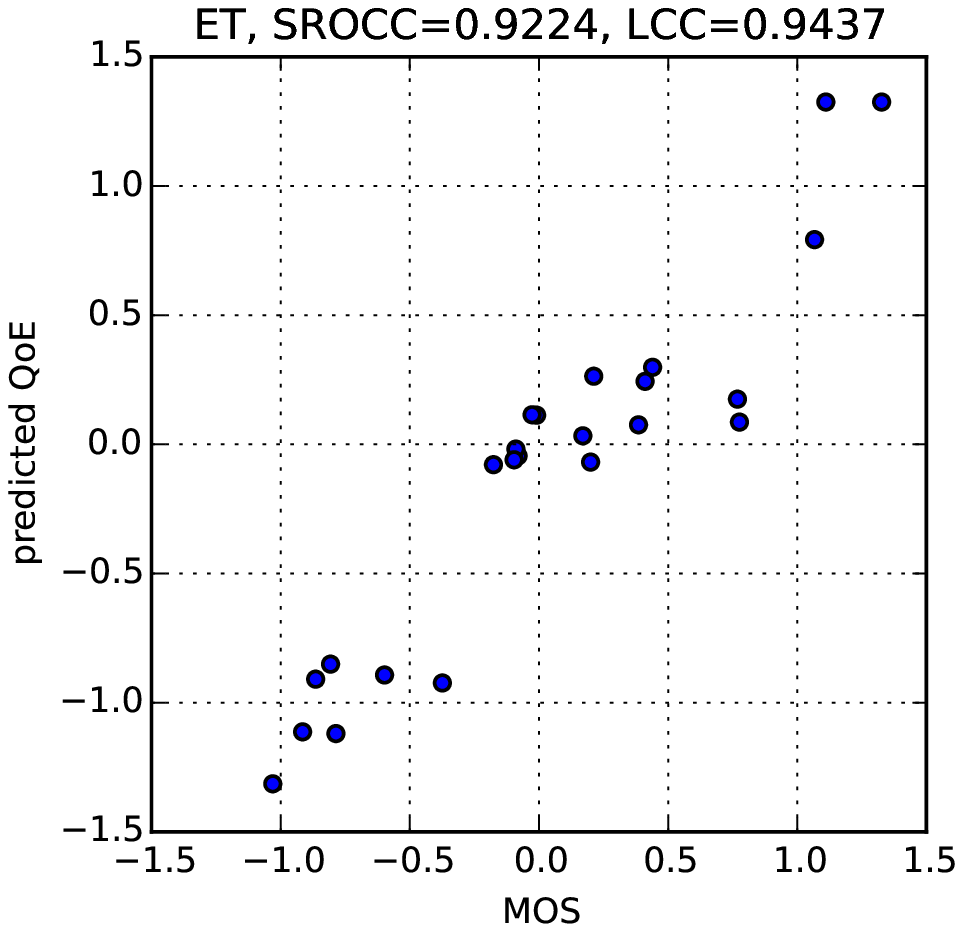}
\includegraphics[width=0.5\columnwidth] {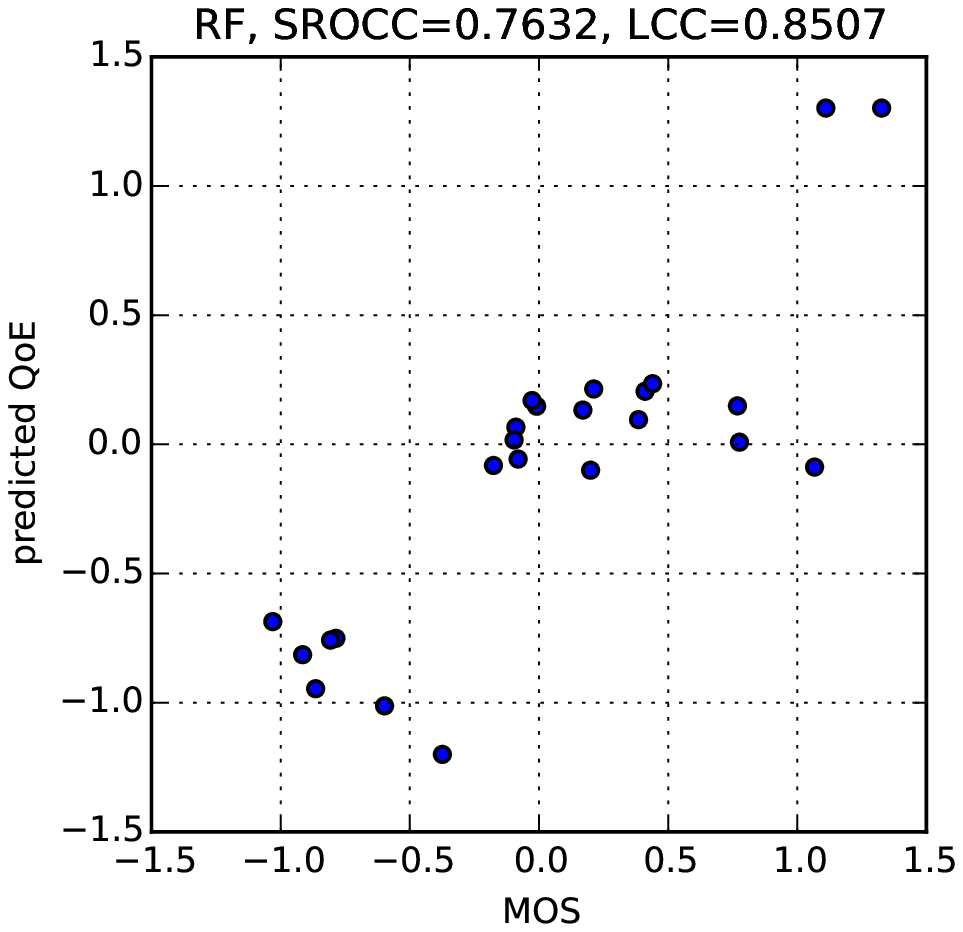}
\includegraphics[width=0.5\columnwidth] {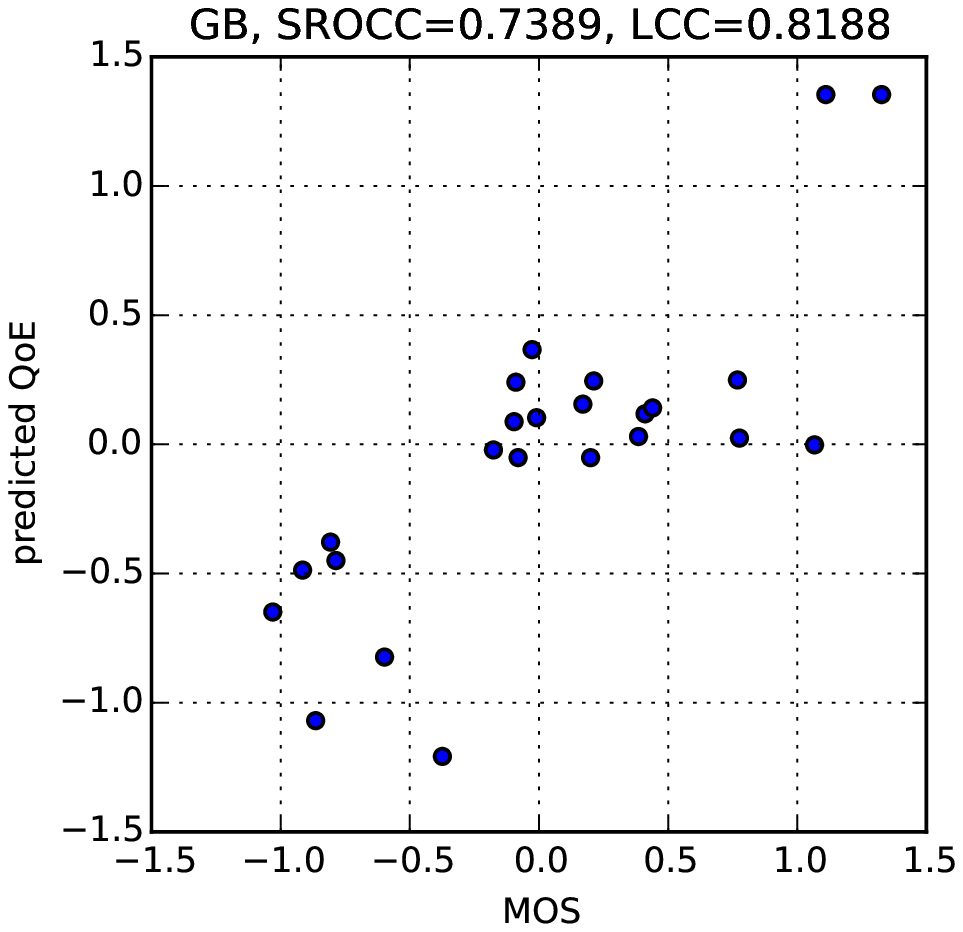}
}
\caption{Predicted QoE scores (horizontal axis) against MOS scores (vertical axis) on a test set when using MS-SSIM across different QoE prediction models. First row: MS-SSIM, MS-SSIM+SQI, MS-SSIM+Ridge, MS-SSIM+Lasso; Second row: MS-SSIM+SVR, MS-SSIM+ET, MS-SSIM+RF, MS-SSIM+GB. All models used the same train/test combination.}
\label{visual_example_all}
\end{figure*}

Finally, we compared models within each of three QoE prediction categories: QoS (FTW and VsQM), VQA-based ones (PSNR, SSIM, MS-SSIM) and hybrid ones (SQI and the proposed Video ATLAS). Table \ref{Compare_LN} shows the median SROCC and LCC results for these methods. A statistical significance test (Wilcoxon ranksum test \cite{siegel1956nonparametric} with significance level $\alpha=0.01$) was carried out by comparing the distributions of SROCC across all $1000$ trials. The results of this analysis are tabulated in Table \ref{stat_sign}. Clearly, Video ATLAS outperformed the other QoE prediction models when using SSIM and MS-SSIM. These improvements are also visually demonstrated in Fig. \ref{visual_example_all} using MS-SSIM for all regression models. In this example, the best performing regression model was ET.

\begin{table}[!ht]
\caption{Results on the LIVE-Netflix DB over $1000$ pre-generated $80\%$ train and $20\%$ test splits. The best result is denoted with bold. For Video ATLAS, we show the best predictor in terms of SROCC and its corresponding LCC.} 
\label{Compare_LN}
\centering
\begin{tabular}{|c|c|c|c|}
\hline
Method & SROCC & LCC & Best \\
\hline
FTW \cite{hossfeld2011quantification} & 0.3403 & 0.2956 & - \\
\hline
VsQM \cite{rodriguez2012quality} & 0.3120 & 0.2421 & - \\
\hline
PSNR & 0.6074 & 0.6048 & - \\
\hline
SSIM \cite{wang2004image} & 0.6748 & 0.7289 & - \\
\hline
MS-SSIM \cite{wang2003multiscale} & 0.6557 & 0.7104 & - \\
\hline
PSNR+SQI \cite{duanmu2016sqi} & 0.6565 & 0.6599 & - \\
\hline
SSIM+SQI \cite{duanmu2016sqi} & 0.7565 & 0.8031 & - \\
\hline
MS-SSIM+SQI \cite{duanmu2016sqi} & 0.7270 & 0.7731 & - \\
\hline
PSNR+ATLAS & 0.6687 & 0.8145 & Ridge \\
\hline
SSIM+ATLAS & 0.8547 & 0.9186 & ET \\
\hline
MS-SSIM+ATLAS & \textbf{0.8752} & \textbf{0.9289} & ET \\
\hline
\end{tabular}
\end{table}

\begin{table*}
\centering
\caption{Statistical significance analysis on the LIVE-Netflix Video QoE Database. Each entry corresponds to the result of a ranksum test between the performances of the methods in the corresponding row and the column. A value of `1' indicates that the row is statistically better than the column, while a value of `0' indicates that the row is statistically worse than the column; a value of `-' indicates that the row and column are indistinguishable. For SQI, we determined the best set of parameters for each train/test split (and quality metric), then used those parameters to compute and analyze SQI. For Video ATLAS, we report only the best performing regressor.}
\label{stat_sign}
\scalebox{0.945}
{
\begin{tabular}{|c|c|c|c|c|c|c|c|c|c|c|c|c|} \hline
\multicolumn{2}{|c|}{ \multirow{2}{*}{} } & \multicolumn{2}{c|}
{QoS} & \multicolumn{3}{c|}
{VQA} & \multicolumn{3}{c|}
{SQI} & \multicolumn{3}{c|}
{ATLAS} \\ \cline{3-13} \multicolumn{2}{|c|}{} & FTW & VsQM & PSNR & SSIM & MS-SSIM & PSNR & SSIM & MS-SSIM & PSNR & SSIM & MS-SSIM \\ \hline
\multirow{2}{*}{QoS} & FTW & -1 & 1 & 0 & 0 & 0 & 0 & 0 & 0 & 0 & 0 & 0 \\ \cline{2-13} & VsQM & 0 & -1 & 0 & 0 & 0 & 0 & 0 & 0 & 0 & 0 & 0 \\ \hline
\multirow{3}{*}{VQA} & PSNR & 1 & 1 & -1 & 0 & 0 & 0 & 0 & 0 & 0 & 0 & 0 \\ \cline{2-13} & SSIM & 1 & 1 & 1 & -1 & -1 & 1 & 0 & 0 & -1 & 0 & 0 \\ \cline{2-13} & MSSIM & 1 & 1 & 1 & -1 & -1 & -1 & 0 & 0 & -1 & 0 & 0 \\ \hline
\multirow{3}{*}{SQI} & PSNR & 1 & 1 & 1 & 0 & -1 & -1 & 0 & 0 & 0 & 0 & 0 \\ \cline{2-13} & SSIM & 1 & 1 & 1 & 1 & 1 & 1 & -1 & 1 & 1 & 0 & 0 \\ \cline{2-13} & MSSIM & 1 & 1 & 1 & 1 & 1 & 1 & 0 & -1 & 1 & 0 & 0 \\ \hline
\multirow{3}{*}{ATLAS} & PSNR & 1 & 1 & 1 & -1 & -1 & 1 & 0 & 0 & -1 & 0 & 0 \\ \cline{2-13} & SSIM & 1 & 1 & 1 & 1 & 1 & 1 & 1 & 1 & 1 & -1 & 0 \\ \cline{2-13} & MSSIM & 1 & 1 & 1 & 1 & 1 & 1 & 1 & 1 & 1 & 1 & -1 \\ \hline
\end{tabular}
}
\end{table*}

\subsubsection{Experiment 2: Testing for Pattern Independence}
We then examined pattern independence when applying Video ATLAS, with results shown in Table \ref{Experiment_2_all_Feats}. Clearly, all of the video quality models were improved by using either SQI and/or VideoATLAS. When combined with either SSIM or MS-SSIM, Video ATLAS improved prediction performance more than SQI did. However, unlike our finding in Experiment 1, we found that not all of the features contributed to the QoE prediction result. To further illuminate this claim, Table \ref{Experiment_2_subset_Feats} tabulates the QoE prediction results on the best feature subset of each regressor when STRRED (with mean pooling) was applied. It may be observed that the combination of the VQA and M features was important for all regressors. This again demonstrates the strong recency/memory effects that contribute to retrospective QoE evaluation. In the case of LCC, Video ATLAS was further improved by including the rebuffering features R$_1$ and/or R$_2$.

\begin{table}[!ht]
\caption{Experiment 2: Results on the LIVE-Netflix DB (video quality metrics, SQI and Video ATLAS). The best result is denoted with bold. For Video ATLAS, we show the best predictor in terms of SROCC and its corresponding LCC. All 5 features are used as input.} 
\label{Experiment_2_all_Feats}
\centering
\begin{tabular}{|c|c|c|c|}
\hline
Method & SROCC & LCC & Best \\
\hline
PSNR & 0.4945 & 0.5312 & - \\
\hline
PSNR+SQI \cite{duanmu2016sqi} & 0.4989 & 0.5340 & - \\
\hline
PSNR+ATLAS & 0.4945 & 0.5321 & Ridge \\
\hline
SSIM \cite{wang2004image} & 0.6615 & 0.7947 & - \\
\hline
SSIM+SQI \cite{duanmu2016sqi} & 0.6791 & 0.7927 & - \\
\hline
SSIM+ATLAS & 0.7143 & 0.8650 & RF \\
\hline
MS-SSIM \cite{wang2003multiscale} & 0.6659 & 0.7982 & - \\
\hline
MS-SSIM+SQI \cite{duanmu2016sqi} & 0.6835 & 0.7955 & - \\
\hline
MS-SSIM+ATLAS & 0.6961 & 0.8345 & GB \\
\hline
NIQE & 0.4681 & 0.4107 & - \\
\hline
NIQE+ATLAS & 0.6447 & 0.6541 & RF \\
\hline
VMAF & 0.3890 & 0.4486 & - \\
\hline
VMAF+ATLAS & 0.7415 & 0.7075 & RF \\
\hline
STRRED & 0.8066 & 0.7848 & - \\
\hline
STRRED+ATLAS & \textbf{0.8198} & \textbf{0.7923} & Ridge \\
\hline
GMSD & 0.4989 & 0.5545 & - \\
\hline
GMSD+ATLAS & 0.5256 & 0.6679 & RF \\
\hline
\end{tabular}
\end{table}

\begin{table*}[!ht]
\caption{Experiment 2: Results on the LIVE-Netflix DB across different feature subsets when the video quality metric is STRRED and mean pooling is applied. The best result is denoted with bold. For each regressor, we show the feature subset resulting in the best SROCC and the feature subset resulting in the best LCC.} 
\label{Experiment_2_subset_Feats}
\centering
\begin{tabular}{|c|c|c|c|c|}
\hline
Regressor & Best SROCC & Feature Set & Best LCC & Feature Set \\
\hline
Ridge & 0.7934 & VQA+M & 0.7939 & VQA+M+R$_1$ \\
\hline
Lasso & 0.7934 & VQA+M & 0.7918 & VQA+M+R$_1$ \\
\hline
SVR & \textbf{0.8242} & VQA+M & 0.8618 & VQA+M+R$_2$ \\
\hline
RF & 0.7385 & VQA+M & 0.8771 & VQA+I+R$_1$+R$_2$ \\
\hline
ET & 0.7437 & VQA+I+R$_1$+R$_2$ & 0.8698 & VQA+M+R$_1$+R$_2$ \\
\hline
GB & 0.8079 & VQA+M+R$_2$ & \textbf{0.8821} & VQA+M+R$_1$ \\
\hline
\end{tabular}
\end{table*}

\subsection{Experiments on the Waterloo Video QoE Database}
The proposed framework uses subjective data to make QoE predictions; hence its predictive power must also be carefully evaluated on other video QoE databases to understand its generalizability. The only publicly available video QoE database that considers other interactions between rebuffering and quality changes is the Waterloo Video QoE Database \cite{duanmu2016sqi} (Waterloo DB). This recently developed database consists of 20 RAW HD 10 sec. reference videos. Each video was encoded using H.264 into three bitrate levels (500Kbps, 1500Kbps, 3000Kbps) yielding 60 compressed videos. For each one of those sequences, two more categories of video sequences were created by simulating a 5 sec. rebuffering event either at the beginning or at the middle of the video sequence. In total, 200 video sequences were evaluated by more than 25 subjects. Based on the collected subjective data, the authors designed the Streaming QoE Index (SQI) to ``account for the instantaneous quality degradation due to perceptual video presentation impairment, the playback stalling events, and the instantaneous interactions between them''. 

Unlike the LIVE-Netflix database (LIVE-Netflix DB), Waterloo DB  consists of short video sequences (which may not reflect the experiences of viewers watching minutes or hours of video content), used fewer subjects, and importantly, the rebuffering events and the bitrate/quality changes were not driven by any realistic assumptions on the available network or the buffer size. However, given its simplicity and the lack of availability of other public domain databases of this type, applying our proposed model framework on this database may yield a comparison of practical worth. We compared the predictive power of our model with SQI \cite{duanmu2016sqi}, FTW \cite{hossfeld2011quantification}, VsQM \cite{rodriguez2012quality} and several VQA models. Aside from SQI and Video ATLAS, the other methods do not consider both rebuffering events and bitrate variations. When conducting direct comparisons, we used only the quality prediction models that were reported for SQI: PSNR, SSIM, MS-SSIM and SSIMplus \cite{rehman2015display}. Given the simple playout patterns, only the VQA+M+R$_2$ feature set was applicable for Video ATLAS. Since the videos in the Waterloo DB do not suffer from dynamic rate changes, the M feature was computed here as the amount of time since a rebuffering event took place. We refer to this feature as M$_{\mathrm{stall}}$. As before, we conducted $1000$ trials, split the contents into training and testing subsets to avoid content bias, and used a pre-generated matrix of such indices. We carried out the following three experiments:

\textit{Experiment 3:} We conducted 1000 trials of 80\% train, 20\% test splits on the Waterloo DB. The results are tabulated in Table \ref{E1}. For Video ATLAS, only the best regression model (in terms of SROCC) is reported. To ensure that SQI yielded its best results on this dataset, we used the parameters suggested in \cite{duanmu2016sqi} (different for each quality model). As before, video quality models did not perform as well as the SQI and Video ATLAS variants. Notably, the performance of MS-SSIM and SSIMplus were worse than that of SSIM even though both have been shown to yield better results than SSIM on the IQA and VQA problems. This verifies our earlier observation: the Waterloo DB contains both rebuffering events and quality changes; hence a better IQA/VQA model may not always correlate better with subjective QoE. Overall, Video ATLAS performed slightly better than SQI, likely in part since the playout patterns in that dataset are simpler, the feature variation is smaller and the number of input features was reduced to only three. Given that SQI was designed on the Waterloo DB, the Video ATLAS results are quite promising.

\begin{table}[!ht]
\caption{Experiment 3: results on the Waterloo DB over $1000$ pre-generated $80\%$ train and $20\%$ test splits. The best result is denoted with bold. For Video ATLAS, we show the best predictor in terms of SROCC and its corresponding LCC.} 
\label{E1}
\centering
\begin{tabular}{|c|c|c|c|}
\hline
Method & SROCC & LCC & Best \\
\hline
FTW \cite{hossfeld2011quantification} & 0.3290 & 0.3358 & - \\
\hline
VsQM \cite{rodriguez2012quality} & 0.2358 & 0.3324 & - \\
\hline
PSNR & 0.6894 & 0.6875 & - \\
\hline
SSIM \cite{wang2004image} & 0.8172 & 0.8544 & - \\
\hline
MS-SSIM \cite{wang2003multiscale} & 0.7986 & 0.8345 & - \\
\hline
SSIMplus \cite{rehman2015display} & 0.8025 & 0.8414 & - \\
\hline
PSNR+SQI \cite{duanmu2016sqi}  & 0.7800 & 0.7535 & - \\
\hline
SSIM+SQI \cite{duanmu2016sqi}  & 0.9085 & 0.9028 & - \\
\hline
MS-SSIM+SQI \cite{duanmu2016sqi} & 0.8891 & 0.8808 & - \\
\hline
SSIMplus+SQI \cite{duanmu2016sqi} & 0.9103 & 0.9012 & - \\
\hline
PSNR+ATLAS & 0.7799 & 0.7510 & SVR \\
\hline
SSIM+ATLAS & \textbf{0.9142} & \textbf{0.9097} & SVR \\
\hline
MS-SSIM+ATLAS & 0.8955 & 0.8880 & Lasso \\
\hline
SSIMplus+ATLAS & 0.9084 & 0.8981 & Ridge \\
\hline
\end{tabular}
\end{table}
Next, we studied the performance of our proposed QoE prediction framework when one of the databases is used for testing and the other for training. In this case, we applied 10-fold cross validation on the entire training dataset to determine the parameters of each regressor. Some regressors, such as RF, may give different results each time; hence we conducted 50 iterations and tabulated the median results in Table \ref{train_one_test_other}. 

\textit{Experiment 4:} We used the Waterloo DB for training and tested the trained models on the LIVE-Netflix DB. For SQI, since we trained on the Waterloo DB, we again used the suggested optimal parameters from \cite{duanmu2016sqi}. For Video ATLAS, we used the Waterloo DB to determine the best parameters for each regressor. The best QoE predictor was Video ATLAS, when combined with SSIM. Clearly, simple QoE predictors based on rebuffering information only such as FTW (or VsQM), or that only uses standard video quality models, perform worse than more general QoE models such as SQI and Video ATLAS. It may also be observed that Video ATLAS outperformed SQI in terms of SROCC and LCC. While Video ATLAS performed better, it should be noted that it used only 3 of the 5 input features (VQA+M$_{\mathrm{stall}}$+R$_2$), given the simple design of the Waterloo DB. A more general dataset for training could potentially increase the predictive performance of Video ATLAS even further.

\textit{Experiment 5:} We then used the LIVE-Netflix DB to train the QoE prediction models, and tested them on the Waterloo DB. Again, to ensure that SQI would yield the best possible results when testing on the Waterloo DB, we used the parameters suggested in \cite{duanmu2016sqi}. For Video ATLAS, we used the Waterloo DB to determine the best parameters of each regressor. As is also shown in Table \ref{train_one_test_other}, SQI and Video ATLAS delivered similar results (Video ATLAS is slightly better when combined with SSIM and MS-SSIM) while FTW, VsQM and objective VQA models performed poorly. Again, when testing on the Waterloo DB, Video ATLAS uses only 3 features, thereby hampering its predictive power. As shown before, combining multiple complimentary features into the Video ATLAS engine is important if it is to achieve its most competitive performance. However, Video ATLAS still competed well against SQI (which was designed and optimized into the Waterloo DB) despite the fact that it was trained on the LIVE-Netflix dataset. This strongly suggests that it generalizes well. By contrast, the results of SQI in experiments 4 and 5 show that it did not generalize as well on the LIVE-Netflix DB.

In experiments 3, 4 and 5, we found that simple learning models such as SVR, Ridge and Lasso, when combined with the three most important features: VQA, M (or M$_{\mathrm{stall}}$) and R$_2$, performed better than SQI and tree-based regressors. This simplicity of Video ATLAS is highly desirable: simple regressors with features that capture the three main properties of subjective QoE (video quality, rebuffering and memory) are more explainable and less likely to overfit on unseen test data.

\begin{table*}[htp]
\caption{Results on the Waterloo DB and the LIVE-Netflix DB when one is used for training and the other for testing. Left: training on the Waterloo DB and testing on the LIVE-Netflix DB; Right: training on LIVE-Netflix and testing on Waterloo. The best result is denoted with bold. For Video ATLAS, we show the best predictor in terms of SROCC and its corresponding LCC.}

\centering
\begin{tabular}{|c|c|c|c|}
\hline
Method & SROCC & LCC & Best \\
\hline
FTW \cite{hossfeld2011quantification} & 0.3352 & 0.2900 & - \\
\hline
VsQM \cite{rodriguez2012quality} & 0.3236 & 0.2374 & - \\
\hline
PSNR & 0.5152 & 0.5073 & - \\
\hline
SSIM \cite{wang2004image} & 0.7015 & 0.7219 & - \\
\hline
MS-SSIM \cite{wang2003multiscale} & 0.6800 & 0.7104 & - \\
\hline
PSNR+SQI \cite{duanmu2016sqi}  & 0.5904 & 0.5905 & - \\
\hline
SSIM+SQI \cite{duanmu2016sqi}  & 0.7451 & 0.7070 & - \\
\hline
MS-SSIM+SQI \cite{duanmu2016sqi} & 0.7239 & 0.6848 & - \\
\hline
PSNR+ATLAS & 0.6155 & 0.6116 & SVR \\
\hline
SSIM+ATLAS & \textbf{0.8203} & \textbf{0.7813} & Lasso \\
\hline
MS-SSIM+ATLAS & 0.8000 & 0.7670 & Lasso \\
\hline
\end{tabular}
\begin{tabular}{|c|c|c|c|}
\hline
Method & SROCC & LCC & Best \\
\hline
FTW \cite{hossfeld2011quantification} & 0.3154 & 0.3313 & - \\
\hline
VsQM \cite{rodriguez2012quality} & 0.2259 & 0.3233 & - \\
\hline
PSNR & 0.6715 & 0.6587 & - \\
\hline
SSIM \cite{wang2004image} & 0.8177 & 0.8408 & - \\
\hline
MS-SSIM \cite{wang2003multiscale} & 0.7928 & 0.8168 & - \\
\hline
PSNR+SQI \cite{duanmu2016sqi}  & 0.7492 & 0.7316 & - \\
\hline
SSIM+SQI \cite{duanmu2016sqi}  & 0.9009 & 0.8897 & - \\
\hline
MS-SSIM+SQI \cite{duanmu2016sqi} & 0.8807 & 0.8652 & - \\
\hline
PSNR+ATLAS & 0.7439 & 0.7254 & SVR \\
\hline
SSIM+ATLAS & \textbf{0.9090} & \textbf{0.8963} & Lasso \\
\hline
MS-SSIM+ATLAS & 0.8888 & 0.8716 & Lasso \\
\hline
\end{tabular}
\label{train_one_test_other}
\end{table*}

\section{Future Work}
\label{the_end}

We described a learning-based approach for QoE prediction that integrates video quality models, rebuffering-aware, and memory features into a single QoE prediction model. 
This framework embodies our first attempt to develop an integrated QoE model, where rebuffering events and quality changes are considered in a unified way. We envision developing more sophisticated models for QoE prediction which could be directly used for continuous time QoE monitoring \cite{chen2014modeling}. 
Towards predicting continuous time scores, combining frame-based objective quality models with temporally varying rebuffering statistics will require a better understanding of how QoE is affected by and further modulated by both inherent short and long term memory effects. 

Towards achieving this goal, time series models such as ARIMA \cite{mills1991time} can be exploited. The LIVE-Netflix Video QoE Database includes continuous time subjective data which is rich and suitable for designing such continuous time QoE models. Therefore, a natural step forward is to deploy prediction methods which also integrate temporal aspects of user QoE in order to design better strategies for the resource allocation problem. However, this remains a challenging problem.

\section{Acknowledgement}The authors would like to acknowledge Zhi Li for his valuable comments on the manuscript. Also, Anush K. Moorthy, Ioannis Katsavounidis and Anne Aaron for their help in designing the LIVE-Netflix Video QoE Database and for sharing their insights on video streaming problems.

\bibliographystyle{IEEEtran}
\bibliography{bibfile}{}

\end{document}